\documentclass{article}

\usepackage{epsfig}
\usepackage{graphicx}
\usepackage{psfig}
\usepackage{color}

\usepackage{amsmath}
\usepackage{amssymb}
\usepackage{cite}
\usepackage{array}
\usepackage{theorem}

\usepackage{textcomp}

\usepackage{fancyhdr}
\newcommand{\beqs}{\begin{equation*}}
\newcommand{\beq}{\begin{equation}}

\newcommand{\eeqs}{\end{equation*}}
\newcommand{\eeq}{\end{equation}}

\newcommand{\beqas}{\begin{eqnarray*}}
\newcommand{\beqa}{\begin{eqnarray}}

\newcommand{\eeqas}{\end{eqnarray*}}
\newcommand{\eeqa}{\end{eqnarray}}

\newcommand{\eps}{\varepsilon}
\newcommand{\al}{\alpha}
\newcommand{\be}{\beta}
\newcommand{\ga}{\gamma}
\newcommand{\de}{\delta}
\newcommand{\om}{\omega}
\newcommand{\ka}{\kappa}
\newcommand{\la}{\lambda}

\newcommand{\Om}{\Omega}

\newcommand{\blist}{\begin{itemize}}

\newcommand{\elist}{\end{itemize}}

\providecommand{\href}[2]{#2}

\newcommand{\unity}{\mathbbm{1}}

\DeclareFontFamily{OT1}{rsfs}{}
\DeclareFontShape{OT1}{rsfs}{m}{n}{ <-7> rsfs5 <7-10> rsfs7 <10->rsfs10}{} 
\DeclareMathAlphabet{\mycal}{OT1}{rsfs}{m}{n}

\newcommand{\sph}{S^2}

\DeclareMathAlphabet{\matheurm}{U}{eur}{m}{n}  
\DeclareMathAlphabet{\matheubf}{U}{eur}{b}{n}  
\usepackage[mathscr]{eucal}                    

\usepackage{xspace} 

\newcommand{\1}{1\hspace{-0.243em}\text{l}} 

\newcommand{\Hodge}{\ast\negthickspace} 
\newcommand{\D}{D} 

\DeclareMathAlphabet{\matheurm}{U}{eur}{m}{n}  

\usepackage{latexsym}
\usepackage[T1]{fontenc}
\newcommand{\thorn}{\mbox{\th}} 

\usepackage{bbm}

\newcommand{\I}{I}
\newcommand{\II}{I\negmedspace I}
\newcommand{\III}{\I\negmedspace,\II}
\newcommand{\lie}{\mathcal{L}}
\newcommand{\killing}{\mathcal{L}}
\newcommand{\bs}[1]{\boldsymbol{#1}}
\newcommand{\e}{\varepsilon}
\newcommand{\dif}{\mathrm{d}}
\newcommand{\re}{\mathrm{Re}}
\newcommand{\im}{\mathrm{Im}}
\newcommand{\lsl}{l\llap{/}}
\newcommand{\nsl}{\rlap{/}n}
\newcommand{\msl}{\rlap{/}m}
\newcommand{\mbsl}{\rlap{/}\bar{m}}

\newcommand{\bra}[1]{\langle #1|}
\newcommand{\ket}[1]{|#1\rangle}
\newcommand{\braket}[2]{\langle #1|#2\rangle}

\newtheorem{definition}{Definition}
\theoremstyle{break}
{\theoremheaderfont{\rmfamily}\theorembodyfont{\rmfamily}\newtheorem{example}{Example}[definition]}

\newcommand{\contra}{\lrcorner}

\begin{document}

\renewcommand{\thefootnote}{\fnsymbol{footnote}}
\thispagestyle{empty}
\begin{titlepage}

\renewcommand{\thefootnote}{\fnsymbol{footnote}}

\vspace*{-2truecm}

\hfill TUW--04--32 \\\mbox{}
\hfill LU-ITP 2004/048


\begin{center}

{\Large\bf The Spherically Symmetric\\ Standard Model with Gravity}
\vspace{1.0cm}

{\bf H.\ Balasin\footnotemark[1], C.\ G.\ B\"ohmer\footnotemark[2] and 
D.\ Grumiller\footnotemark[3]}
\vspace{5ex}

\footnotemark[1]\footnotemark[2]\footnotemark[3]{Institut f\"ur
    Theoretische Physik \\ Technische Universit\"at Wien \\ Wiedner
    Hauptstr.  8--10, A-1040 Wien \\ Austria}
\vspace{2ex}

\footnotemark[3]{Institut f\"ur
    Theoretische Physik \\ Universit\"at Leipzig \\ Augustusplatz 10-11,
    D-04103 Leipzig \\ Germany
}

\vspace{1cm}

\footnotetext[1]{e-mail: {\tt hbalasin@tph.tuwien.ac.at}}
\footnotetext[2]{e-mail: {\tt boehmer@hep.itp.tuwien.ac.at}}
\footnotetext[3]{e-mail: {\tt grumil@hep.itp.tuwien.ac.at}}

\end{center}

\begin{abstract}
Spherical reduction of generic four-dimensional theories is revisited. Three 
different notions of ``spherical symmetry'' are defined. The following 
sectors are investigated: Einstein-Cartan theory, 
spi\-nors, (non-)abelian gauge fields and scalar 
fields. In each sector a different formalism seems to be most convenient: 
the Cartan formulation of gravity works best in the purely gravitational 
sector, the Einstein formulation is convenient for the Yang-Mills sector 
and for reducing scalar fields, and the Newman-Penrose formalism seems to 
be the most transparent one in the fermionic sector. Combining them 
the spherically reduced Standard Model of particle physics 
together with the usually omitted gravity part can be presented as 
a two-dimensional (dilaton gravity) theory. 

%

\end{abstract}


\vspace{2ex}
\noindent
{\em Key words: spherical symmetry, Standard Model, $2d$ theories, dilaton gravity}

\vfill
\end{titlepage}


\renewcommand{\thefootnote}{\arabic{footnote}}
\setcounter{footnote}{0}
\numberwithin{equation}{section}

\tableofcontents

\listoffigures

\thispagestyle{fancy}
\newpage

\section{Introduction}

Spherical symmetry plays a pivotal role in theoretical physics 
because the system simplifies such that an exact solution is 
often possible; this in turn allows for an understanding of some 
basic principles of the underlying dynamical system and 
thus can be of considerable pedagogical value.\footnote{It is impossible to present a complete list of references 
regarding spherical symmetry, because ever since Coulomb we estimate that ca.~$10^5$ publications appeared in this context. 
However, whenever a certain technical detail is used of course some of the original literature, or at least reviews for 
further orientation, will be provided.}

As compared to other frequently used scenarios, like the
ultra-relativistic limit where the rest mass is much smaller than 
the kinetic energy or the static limit where the rest mass is much larger 
than the kinetic energy, spherical symmetry has the advantage that it allows 
for dynamics such as scattering of s-waves as opposed to the static case 
{\em and} for bound states as opposed to the ultra-relativistic limit.
Moreover, many physical systems of relevance exhibit at least 
approximate spherical symmetry -- to name a few: the $l=0$ 
sector of the Hydrogen atom, non-rotating isolated stars, 
the universe on large scales (actually isotropic with respect to 
any point), etc. Also semi-classically 
spherically symmetric modes are often the dominant ones -- 
e.g. ca. 90\% of the Hawking flux of an evaporating black hole is
due to this sector (cf.~e.g.~\cite{Frolov:1998wf}), the 
Balmer series stems from it (disregarding the finestructure), etc. 
From a technical point of view the success of spherical symmetry is
related to the fact that systems in two dimensions $d=2$ have many 
favourable properties (cf.~e.g.~\cite{Efthimiou:2000gz}).

However, the advantages of simplifications due to spherical symmetry 
become most apparent in the context of (quantum) gravity.
As an illustration four selected examples are presented: Krasnov and 
Solodukhin discussed recently the wave equation on
spherically symmetric black hole (BH) backgrounds~\cite{Krasnov:2004ki}. 
They found an intriguing interpretation in terms of
Conformal Field Theory, at least in certain limits
(near horizon, near singularity and high damping), thus realizing 
'tHooft's suggestion \cite{'tHooft:1990fr} of an analogy 
between strings and BHs. 
In the framework of canonical quantum gravity recently the concepts of quantum horizons \cite{Bojowald:2004si} and quantum black holes \cite{Husain:2004yy} have been introduced for spherically symmetric systems. While the former work is inspired by the concept of isolated horizons \cite{Ashtekar:2000sz}, the latter invokes trapped surfaces and thus may be applied to dynamical horizons. Both confirm the heuristic picture that at the quantum level horizons fluctuate.
One of the present authors
 together with Fischer, Kummer and Vassilevich 
considered scattering of s-waves on their own gravitational self-energy by 
means of two-dimensional methods, obtaining a simple 
but nontrivial S-matrix with virtual BHs as intermediate states~\cite{Fischer:2001vz} 
(for a review cf.~\cite{Grumiller:2004yq}), in accordance 
with 'tHooft's idea that BHs have to be considered in the S-matrix together with elementary matter 
fields~\cite{'tHooft:1996tq}. Finally, the seminal 
numerical work by Choptuik \cite{Choptuik:1993jv} on critical 
collapse was based upon a study of the spherically symmetric 
Einstein-massless-Klein-Gordon model. Although similar 
features were found later in many other systems (for a review 
cf.~\cite{Gundlach:2002sx}) we believe it is no coincidence 
that the crucial discovery was made first in the simpler 
spherically symmetric case. 

In the first, third and fourth example the coupling to matter degrees 
of freedom was essential. It is therefore of some interest to study 
the most general coupling to matter consistent with observation, 
in particular the Standard Model of particle physics \cite{Weinberg:1967tq} or a recent improvement thereof \cite{Davoudiasl:2004be}.

The purpose of this work is 1.\ to clarify what is meant by spherical 
symmetry; three different notions will be presented, 2.\ to review the
basic formalisms that are useful in the context of spheric reduction
(Cartan-, Geroch-Held-Penrose (GHP) and metric-formalism), 3.\ to apply them
to obtain the spherically symmetric Standard Model plus
gravity (SSSMG) in a comprehensive manner, 
4.\ to present an effective theory 
in $d=2$ which then in principle can be quantised. As byproducts several 
sectors will be discussed in technical detail. Necessarily, large part 
of this work have the character of a review. Nonetheless, several new 
results are contained in it: It is shown
that static perfect fluids can be regarded as generalised dilaton 
gravity models. In the reduction of the Einstein-Yang-Mills-Dirac system we find an
additional contribution that might have been overlooked in previous
considerations. We discuss the symmetry restoration of spontaneous
symmetry breaking by giving an interpretation to the effective Higgs potential.
The Yukawa interaction is spherically reduced without fixing
the isospin direction. We spherically reduce the torsion induced four fermion
interaction term present in Einstein-Cartan theory. Finally we
comment on the quantisation of the SSSMG. 

We would also like to point out that one of the main aims of our 
work is to provide a link between concepts from particle physics 
(like the matter content inspired by the standard model) and 
input from general relativity like the spin-coefficient formalism 
that is particularly adequate for the reduction of 
spinors in a spherically symmetric context. 

This paper is organised as follows: in Section~\ref{se:2} three
different notions of spherical symmetry are discussed.
Section~\ref{se:3} fixes the notation and introduces the three formalisms (Cartan, GHP,
metric) by means of relevant and rather explicit examples. A brief 
recapitulation of dilaton gravity with matter by means of a discussion 
of static perfect fluid solutions is given. Collecting them the spherically 
symmetric Standard Model plus gravity is constructed
in Section~\ref{se:sm}. It is presented as an effective 
theory in two dimensions~(Section~\ref{se:4}). 
 The final Section~\ref{se:con} contains some concluding remarks.
The appendices provide supplementary material 
mostly related to the GHP formalism.

\section{Three ways of spherical symmetry/reduction}
\label{se:2}

In the following we will define different notions of spherical symmetry.\\
\mbox{}\\
0) In order to be able to talk about spherical symmetry one needs
an action of the rotation group $SO(3)$ on the spacetime manifold
$\mathcal{M}$ under consideration. For the geometry 
we require that the vector-fields of the action $\xi$ 
leave the metric  $g_{ab}$ unchanged, i.e.,\footnote{Because from the context it will always be clear whether we mean a Lie-group or its associated algebra we do not discriminate notationally between them.}
\begin{align}
      \lie_{\xi}g_{ab}=\nabla_{(a}\xi_{b)}=0 \,,
      \qquad \xi \in SO(3)\,,
      \label{srg:0}
\end{align}
or equivalently that they are Killing. (Let us remark that the action
is given by space-like vector-fields). This property will be assumed
subsequently. It entails the form of the symmetry generators,
a basis of $SO(3)$, 
and the metric
\begin{align}
      \xi_\phi = i \partial_{\phi}\,,\quad \xi_{\pm} = 
      \frac{1}{\sqrt{2}}e^{\pm i\phi}
      \left(\partial_{\theta}\pm i\cot\negthinspace\theta\partial_{\phi}\right)\,,
      \label{srg:1} \\
      \dif s^2=g_{\al\be}(x^\al)\dif x^\al\otimes \dif x^\be -X(x^\al)\dif\Om^2\,,
      \quad \al,\be=0,1\,,
      \label{srg:2}
\end{align}
in adapted coordinates, where $\phi,\theta$ are the standard coordinates of (the round)
$S^2$, $\dif\Om^2=\dif\theta^2+\sin^2\negthinspace{\theta}\,\dif\phi^2$. 
The Killing vectors (\ref{srg:1}) obey the angular momentum 
algebra $[\xi_{\pm},\xi_\phi]=\pm\xi_\pm$, $[\xi_+,\xi_-]=\xi_\phi$.

The state of the physical systems under consideration will in addition
to the metric also contain various matter fields which we denote by
$\phi_{\alpha}$ and which are taken to be sections of various (vector-)bundles
over spacetime. As long as those bundles are naturally tied to $\mathcal{M}$,
i.e., are tensor products of the tangent- and cotangent bundles $T\mathcal{M}$
and $T^{*}\mathcal{M}$, the action of $\xi$ on their sections
is well-defined.\\
\mbox{}\\
1) For these matter fields \emph{strict spherical symmetry} is defined by
$\lie_{\xi}\phi_{\alpha}=0$.
\addtocounter{definition}{1}
\begin{example}[Reduction of scalar matter I] 
\label{ex_scalar1}
The action for scalar matter in $d=4$ reads
\begin{align}
        L^{(4)} = \int \left( G^{\mu\nu} \nabla_\mu \phi \nabla_\nu \phi \right) \omega_G,
        \label{s_action}
\end{align}
where $\phi$ is the scalar field and $\omega_G$ is the $4d$ volume
form. $\lie_{\xi}\phi=0$ implies that in adapted coordinates $\phi=\phi(t,r)$ and 
hence the action~(\ref{s_action}) after integrating out the 
angular part simply leads to the reduced action with $2d$ volume 
form $\omega_g$
\begin{align}
        L^{(2)}=4\pi\int (g^{\alpha\beta}\nabla_\alpha\phi\nabla_\beta\phi )X\omega_g.
\label{eq:example1}
\end{align} 
\end{example}
However, in general we are also interested in bundles more loosely
tied to the spacetime manifold (like Spin-bundles and principal $SU(N)$-bundles).
For their sections the $SO(3)$ action on $\mathcal{M}$ does not
automatically extend. However, in the above mentioned cases there
exist certain {}``natural'' notions, which allow an action to be
defined on the sections of these bundles. In general strict invariance
will not be possible (or too restrictive).
\begin{example}[No strictly spherically symmetric spinors]
Let $k^A=k^0 o^A + k^1 \iota^A$ be an arbitrary spinor. This
spinor would be called strictly spherically symmetric if one
could solve $\lie_{\xi} k^A=0$ for non-trivial $k^0$ and $k^1$.
Direct calculation easily shows that this is impossible.
\end{example}
2) For these fields
only {}``covariant'' transformation behaviour is possible. They will be called \emph{spherically symmetric} if
\begin{align}
      \lie_{\xi}\phi_{\alpha}=D(\xi)\phi_{\alpha}\,,
\end{align}
where $D$ refers to a typically linear transformation,
e.g. a derivative operator.
\setcounter{example}{0}
\addtocounter{definition}{1}
\begin{example}[Gauge fields]
A gauge field obeying $\lie_{\xi_i} A = DW_i$, where $D$ is the gauge
covariant derivative; thus, the field $A$ itself need not be strictly 
spherically symmetric, only up to gauge transformations.
\end{example}
Finally an even less stringent form, which we call \emph{weak spherical
symmetry}, may be defined by expanding the fields with respect to
a complete set of eigenfunctions of the spherical Laplacian.\\
\mbox{}\\
3) For these fields of spin $s$, we decompose
\begin{align}
      \Delta_{S^{2}}\,_sY_{j,m}=-(j-s)(j+s+1)\,_sY_{j,m}\,,\qquad
      \phi_{\alpha}=\sum_{jm}\phi_{\alpha,jm}\,_sY_{j,m}\,,
\end{align}
where $_sY_{j,m}$ are the spin-weighted spherical harmonics (for $s=0$ they coincide with the standard spherical harmonics while for higher spin we refer the reader to subsection~\ref{subsec_red_fer}).
On the dynamical level, i.e., upon insertion into the action
and integration of the angular part, this yields a spherically reduced
(two-dimensional) system.
\setcounter{example}{0}
\addtocounter{definition}{1}
\begin{example}[Reduction of scalar matter II] 
\label{ex_scalar}
The action for scalar matter reads
\begin{align}
        L = \int \left( G^{\mu\nu} \nabla_\mu \phi \nabla_\nu \phi \right) \omega_G,
        \label{scalaraction}
\end{align}
where $\phi$ is the scalar field. Expanding $\phi$ in terms
of spherical harmonics $\phi = \sum_{lm} \phi_{lm} Y_{lm}$,
the scalar action~(\ref{scalaraction}) upon integration of the angular part leads to
\begin{align}
        L = \sum_{lm} \int \left( g^{\alpha\beta} \nabla_\alpha \phi_{lm} 
        \nabla_\beta \phi_{lm} + \frac{l(l+1)}{X} \phi_{lm} \phi_{lm}\right) 
        X \omega_g,
        \label{scalarred}
\end{align}
where the s-wave sector $l=0$ corresponds to (\ref{eq:example1}). 
\end{example}

\begin{example}[Spherically reduced gravity]
Spherically reduced gravity (SRG) emerges from averaging over the angular part,
\begin{align*}
      <R_{\mu\nu}>-1/2<g_{\mu\nu}R>=\ka<T_{\mu\nu}>\,,
\end{align*} 
where $R_{\mu\nu}$ is the Ricci tensor, $T_{\mu\nu}$ is the energy-momentum tensor, 
$\ka$ is the gravitational coupling and the bracket denotes integration 
over the angular part -- this system of averaged equations of motion can be deduced from an 
action in $d=2$, the geometric part of which is just the spherically reduced Einstein-Hilbert action.
For the Einstein-massless-Klein-Gordon model the matter 
part (\ref{scalarred}) contains an infinite tower of scalar fields with dilaton 
dependent (and $l$-dependent) mass.
\end{example}
Each of these notions is weaker than its predecessor:
\begin{align*}
      \textbf{strict sph. sym.}\ \geq\ \textbf{spherical symmetry}
      \ \geq\ \textbf{weak sph. sym.}
\end{align*}
For the rest of this paper we assume spherical symmetry according to 
the second notion, unless stated otherwise.


Having defined spherical symmetry we would like to focus on spherical
reduction. By this procedure we mean the derivation of a reduced action
in $d=2$ the equations of motion of which are equivalent (in a
well-defined way) to the equations of motion of the original theory if
the latter are restricted to spherical symmetry. The fact that such a
procedure works is not trivial in general (i.e.,\ if the isometry group
is different from $SO(3)$). Due to the compactness of $SO(3)$, however,
one can immediately apply Theorem 5.17 (or proposition 5.11) of
\cite{Fels:2001rv} and employ the ``principle of symmetric criticality''
\cite{Palais:1979} which guarantees the (classical) equivalence of the
reduced theory to the original one (cf.~also \cite{Deser:2003up}). The main advantage of spherical
reduction is the possibility to exploit the simplicity of two
dimensional field theories.

\section{Three formalisms}\label{se:3}

The purpose of this section is threefold: the three relevant formalisms are 
reviewed together with their respective advantages, relevant examples are 
considered and {\em en passant} our notation is fixed in detail.

\subsection{Cartan's form calculus and Gravity}
\label{subsec_cartan}
In the Cartan formalism one works in an anholonomic frame and
uses the vielbein 1-form and connection 1-form as independent 
variables. With these variables one can use the advantages of the 
form calculus, where diffeomorphism invariance is implied automatically,
see e.g.~\cite{Ivey:2003}.

\subsubsection{The 2-2 split}

In the Cartan formalism the line element can be written as
\begin{align}
      \dif s^2 = g_{\mu \nu}\, \dif x^{\mu} \otimes \dif x^{\nu} =
      e_{\mu}^m e_{\nu}^n \eta_{mn}\, \dif x^{\mu} \otimes \dif x^{\nu} =
      \eta_{mn}\, e^m \otimes e^n\,.
      \label{metric_viel}
\end{align}
Greek letters are used for holonomic indices and Latin letters for 
anholonomic ones. $\eta_{mn}$ is the flat (Minkowski) metric with 
signature $(+,-,-,-)$. The vielbein is denoted by $E_m^{\mu}$,
\begin{align}
      E_m^{\mu} e_{\mu}^n = \delta_m^n\,,\qquad E_m \contra e^n =\delta_m^n\,,
\end{align}
where $\contra$ means contraction. One similarly writes the vector 
field $E_m = E_m^{\mu} \partial_{\mu}$. 
The covariant derivative is written as
\begin{align}
      \tilde{D}^m{}_n = \delta^m_n \dif + \tilde{\omega}^m{}_n\,,
      \label{cov_der_form}
\end{align}
with the skew-symmetric connection 1-from 
$\tilde{\omega}_{mn} = -\tilde{\omega}_{nm}$, 
because of metricity. The connection 1-form may be split
accordingly
\begin{align}
      \tilde{\omega}^m{}_n = \omega^m{}_n + K^m{}_n\,,
      \label{con_form}
\end{align}
where $\omega^m{}_n$ is the torsion free part and 
$K^m{}_n$ is the contortion.
 
Acting with~(\ref{cov_der_form}) on the vielbein $e^m$ 
and on the connection 1-form $\tilde{\omega}^m{}_n$ defines
the torsion 2-form and the curvature 2-form, respectively
\begin{align}
      T^m &= (\tilde{D} e)^m = \dif e^m + \tilde{\omega}^m{}_n  e^n
      = K^m{}_n  e^n
      = \frac{1}{2} T^m{}_{\mu \nu}\, \dif x^{\mu}  \dif x^{\nu}, 
      \label{tor_form} \\
      R^m{}_n &= (\tilde{D}^2)^m{}_n = \dif \tilde{\omega}^m{}_n 
      + \tilde{\omega}^m{}_l  \tilde{\omega}^l{}_n 
      = \frac{1}{2} R^m{}_{n \mu \nu}\, \dif x^{\mu}  \dif x^{\nu}\,.
      \label{cur_form}
\end{align}
Note that we avoid writing out the wedge product explicitly.

In case of spherical symmetry one can separate the 
metric~(\ref{metric_viel}) 
\begin{align}
      \dif s^2 = \eta_{mn}\, e^m \otimes e^n = \eta_{ab}\, e^a \otimes e^b 
      -\delta_{rs}\, e^s \otimes e^r\,,
\end{align}
where the indices $(\alpha,\beta,\ldots;a,b,\ldots)$ denote quantities of the 
two-dimensional manifold $L$ and letters $(\rho,\sigma,\ldots;r,s,\ldots)$ 
quantities connected with the sphere $\sph$.
Moreover the dilaton $X$ from~(\ref{srg:2}) has been redefined 
as $X = \Phi^2$ in order to avoid square-roots in subsequent 
formulas. Barred (``intrinsic'') and unbarred quantities 
are related by 
\begin{alignat}{2}
      e^a &= \bar{e}^a,&\qquad e^r &= \Phi \bar{e}^r\,, \nonumber \\
      E_a &= \bar{E}_a,&\qquad E_r &= \frac{1}{\Phi} \bar{E}_r\,.
\end{alignat}
The torsion free connection 1-form $\omega^m{}_n$ 
is given by
\begin{alignat}{2}
      \omega^a{}_b &= \bar{\omega}^a{}_b,&\qquad 
      \omega^a{}_r &= (\bar{E}^a \Phi) \bar{e}_r, \nonumber \\
      \omega^r{}_s &= \bar{\omega}^r{}_s,&\qquad 
      \omega^r{}_a &= (\bar{E}_a \Phi) \bar{e}^r. 
      \label{omega_form}
\end{alignat}

\subsubsection{Reduction of torsion and curvature}

Gauge covariant transformation behaviour under spherical symmetry restricts 
the possible contributions of torsion~\cite{Baekler:1980ss} according to 
\begin{align}
      \lie_{\xi} T^a=0\,,\quad\lie_{\xi} T^r=\Om^r{}_sT^s\,,
      \label{srg:15}
\end{align}
where the skew-symmetric matrix $\Om$ is defined by $\lie_{\xi}e^r=\Om^r{}_se^s$. This -- together with the non-existence of a rotationally invariant vector field on $\sph$ -- entails the decomposition
\begin{align}
      T^a=\bar{T}^a+\frac12 T^a{}_{rs}e^re^s\,,\quad T^r=T^r{}_{as}e^ae^s\,.
      \label{srg:17}
\end{align}
Consequently, the sphere $\sph$ has to be torsion free intrinsically, 
i.e.,\ $\bar{E}_t\contra K^r{}_s = 0$.

The connection need not be strictly spherically symmetric but only symmetric up to gauge transformations (much like in the Yang-Mills case below). Expanding (\ref{srg:15}) according to our 2-2 split the conditions read
\begin{align}
      \dif\lie_{\xi}e^a + \lie_{\xi}(\tilde{\om}^a{}_b e^b) + \lie_{\xi}(\tilde{\om}^a{}_r e^r) &= 0\,, \label{srg:10} \\
      \dif\lie_{\xi}e^r + \lie_{\xi}(\tilde{\om}^r{}_ae^a) + \lie_{\xi}(\tilde{\om}^r{}_se^s) &= \Om^r{}_sT^s\,. \label{srg:11}
\end{align}
Because of $\lie_{\xi}e^a=0=\lie_{\xi}\tilde{\om}^a{}_b$ equation~(\ref{srg:10}) establishes
\begin{align}
      \lie_{\xi}K^a{}_r = \tilde{k}^ae_r+\tilde{h}^a\eps_{rs}e^s\,,
      \quad \tilde{k}^a=\tilde{k}^a(x^\al),\,\tilde{h}^a=\tilde{h}^a(x^\al)\,. 
      \label{srg:13}
\end{align}
Plugging~(\ref{con_form}) and~(\ref{omega_form}) into~(\ref{srg:11}) yields
\begin{align}
      \lie_{\xi}K^r{}_a=\Om^r{}_s K^s{}_a\,,
      \label{srg:12}
\end{align} 
This means that the right hand side of (\ref{srg:13}) is not only valid for $\lie_{\xi_i}K^r{}_a$ but also for $K^a{}_r$, which suggests the useful definitions of vector valued scalars
\begin{align}
      k_a:=\bar{E}_r \contra K^r{}_a\,,\qquad 
      h_a:=\bar{E}^s\contra K^r{}_a \eps_{rs}\,.
      \label{srg:14}
\end{align}
The following observation is helpful: The intrinsic torsion $\bar{T}^a$ of the reduced $2d$ theory is irrelevant as there is no way to couple sources to it (because the $2d$ connection is not present in the action of $2d$ fermions). Thus, it can be demanded always $\bar{E}_c\contra K^a{}_b = 0$. Obviously, due to spherical symmetry also $\bar{E}_r\contra K^a{}_b = 0$. Thus, $K^a{}_b$ can be set to zero.

The curvature 2-form~(\ref{cur_form}) can be calculated using~(\ref{con_form}), 
(\ref{omega_form}) and taking into account the previous considerations on torsion. The 2-2 split for curvature and torsion yields the result
\begin{alignat}{3}
R^a{}_b &= \bar{R}^a{}_b+{\mathcal R}^a{}_b\,,&\quad R^r{}_s &= \bar{R}^r{}_s(1+(\bar{E}_a\Phi)(\bar{E}^a\Phi))+{\mathcal R}^r{}_s\,, \\
R^a{}_r &= -\eta^{ab}R_{rb}\,,&\quad R^r{}_a &=(\bar{E}_b\bar{E}_a\Phi)\bar{e}^b\bar{e}^r-(\bar{E}_b\Phi)\om^b{}_a\bar{e}^r+{\mathcal R}^r{}_a\,,\\
T^a &= K^a{}_re^r\,,&\quad T^r &= K^r{}_ae^a+K^r{}_se^s\,, \\
K^a{}_b &= 0\,,&\quad K^r{}_s &= \eps^r{}_s s_a e^a\,, \\
K^a{}_r &= -\eta^{ab}K_{rb}\,,&\quad K^r{}_a &= -\frac12(k_a \bar{e}^r + h_a \eps^{rs} \bar{e}_s)\,,
\end{alignat}
with $\bar{R}^r{}_s=\bar{e}^r\bar{e}_s$ being the curvature 2-form of $\sph$ and $\bar{R}^a{}_b$ being the intrinsic curvature in $2d$. The contortion contributions to curvature read
\begin{align}
{\mathcal R}^r{}_s 
&= \om^r{}_aK^a{}_s+K^r{}_a\om^a{}_s+K^r{}_aK^a{}_s\nonumber \\
&\quad + \dif K^r{}_s +\om^r{}_tK^t{}_s+K^r{}_t\om^t{}_s\,,\\
{\mathcal R}^r{}_a 
&=\dif K^r{}_a+K^r{}_b\om^b{}_a+K^r{}_sK^s{}_a\nonumber\\
&\quad +\eps^r{}_s s_b\bar{e}^b(E_a\Phi)\bar{e}^s-\frac12 \om^r{}_s (k_a \bar{e}^s + h_a \eps^{sr} \bar{e}_r)\,,\\ 
{\mathcal R}^a{}_b 
&= 0 \nonumber\\
&\quad -\frac14 \eta^{ac} (2(\bar{E}_{[c}\Phi) + k_{[c})h_{b]}\bar{e}_r\eps^{rs}\bar{e}_s\,,
\end{align}
where we use $T_{[\mu\nu]}:=T_{\mu\nu}-T_{\nu\mu}$.
Note that in each equation the second line does not contribute to the curvature scalar because the corresponding contractions vanish. Thus, for instance, the contortion contribution ${\mathcal R}^a{}_b$ does not produce any terms in the Einstein-Hilbert action.

As compared to the torsionless case additional effective fields are obtained: three vector valued scalars (depending on $x^\al$), $k_a$, $h_a$ and $s_a$. Depending on the original action in $d=4$ some of these fields might drop. 

\subsubsection{Reduction of the Einstein-Hilbert action}
\label{subsub_hilbert}

Double contraction,  
\begin{align}
      \tilde{R}=R + 2E^a\contra E_r\contra \mathcal{R}^r{}_a 
      + E^s\contra E_r\contra \mathcal{R}^r{}_s\,,
      \label{srg:18}
\end{align}
yields the torsion free curvature scalar $R$ in terms of the 
two-dimensional one $R^L$, terms coming from intrinsic and 
extrinsic curvature of $\sph$ and torsion terms
\begin{multline}
      \tilde{R} = R^L - \frac{2}{\Phi^2} \bigl(1+(\nabla_a\Phi)(\nabla^a\Phi)\bigr)
      -\frac{4}{\Phi}(\nabla_a\nabla^a\Phi)\\
      +\frac{1}{\Phi^2} \frac12 (h_ah^a-k_ak^a) + \frac{2}{\Phi}s_ah^a
      +\frac{2}{\Phi^2}\nabla_a(k^a\Phi)\,.
      \label{ricci_sc}
\end{multline}
The first line coincides with the torsion-free result (e.g. equation~(A.8) of reference~\cite{Grumiller:2002nm}). Note that
in~(\ref{ricci_sc}) $\nabla_a$ is the covariant derivative
operator with respect to $L$. The last term together with the volume form produces just a surface term. Thus, as can be expected on general grounds~\cite{Hehl:1973} 
torsion is not propagating in the Einstein-Hilbert action.

For easier comparability with later results the anholonomic components of
the contortion 1-form $K^m{}_n = K_l{}^m{}_n e^l$ are decomposed into the contortion vector $k_a$ like in (\ref{srg:14}),
\begin{align}   
      k_a = \Phi K_r{}_a{}^r\,,\qquad A^a =\frac{1}{3!}\eps^{almn}K_{lmn}\,,
      \label{vector}
\end{align}
and the axial contortion vector $A^a$. The
remaining components of the contortion tensor are denoted by
$U_{lmn}$. Then the curvature scalar~(\ref{ricci_sc}) can alternatively
be written
\begin{multline}
      \tilde{R} = R^L - \frac{2}{\Phi^2} \bigl(1+(\nabla_a\Phi)(\nabla^a\Phi)\bigr)
      -\frac{4}{\Phi}(\nabla_a\nabla^a\Phi) 
      \\
      -\frac{1}{\Phi^2}\frac{1}{2}k^2 - U^2 -6A^2 +\frac{2}{\Phi^2}\nabla_a(k^a\Phi)\,,
\end{multline}
where 
\begin{align}
      6 A^2 &= -\frac{2}{3}(s_a s^a +\frac{1}{\Phi^2}h_a h^a+\frac{2}{\Phi}s_a h^a )\,,\\
      U^2 &= \frac{2}{3}(s_a s^a +\frac{1}{4\Phi^2}h_a h^a-\frac{1}{\Phi}s_a h^a )\,.
      \label{u2}
\end{align}
This second form of the curvature scalar is used in 
section~\ref{ectheory}. Moreover the separation of the contortion
tensor into its irreducible parts is often found in 
literature~\cite{Alimohammadi:1998vx}. 

In the absence of torsion spheric reduction \cite{Berger:1972pg} 
of the Einstein-Hilbert action $L_{\rm EH}=\int_M R\omega_G$ yields the dilaton gravity action
\begin{align}
      L_{\rm dil}[g_{\alpha\beta},X]=
      4\pi \int_L \left(XR^L+(\nabla X)^2/(2X)-2\right)\om_g\,,
      \label{srg:3}
\end{align}
where $\om_G=\Phi^2\om_g\dif^2\Om$. 
$M$ denotes the four-dimensional manifold and $L$ its
two-dimensional Lorentzian part.

It is convenient to reformulate this second order action\footnote{Alternatively, one can try to eliminate the dilaton by means of its EOM, thus obtaining an action which depends non-linearly on curvature. Reviews on this approach are \cite{Schmidt:1999wb} and \cite{Obukhov:1997uc}.} as a first order one \cite{Schaller:1994es} and to rescale the dilaton as $X\to\la^2 X$ in order to make it dimensionless, 
\begin{align}
      L_{\rm FOG}[e^a,\omega,X,X^a]=\frac{2\pi}{\la^2}\int_L \left[X_a (D\wedge e)^a +X\dif\omega +{\mathcal V} (X,X^aX_a)\epsilon \right]\,,
      \label{cs:1}
\end{align}
with ${\mathcal V}=-X_aX^a/(4X)-\la^2$. Whenever a first order action in
$d=2$ is presented for sake of compatibility with
\cite{Grumiller:2002nm} the following notation is used: in accordance
with above $e^a$ is the zweibein 1-form, $\epsilon = e^+\wedge e^-$ is
the volume 2-form. The 1-form $\omega$ represents the
spin-connection\footnote{It should be noted that even in the absence of
torsion in $d=4$ the connection $\om^a{}_b$ in (\ref{cs:1}) is torsion free if and only if ${\mathcal V}$ depends on $X$ only.} $\om^a{}_b=\eps^a{}_b\om$ with  the totally antisymmetric
Levi-Civit{\'a} symbol $\eps_{ab}$ ($\eps_{01}=+1$). With the flat 
metric $\eta_{ab}$ in light-cone coordinates 
($\eta_{+-}=1=\eta_{-+}$, $\eta_{++}=0=\eta_{--}$) 
the first (``torsion'') term of (\ref{cs:1}) is given 
by $X_a(D\wedge e)^a =\eta_{ab}X^b(D\wedge e)^a =X^+(\dif-\omega)\wedge e^-
+ X^-(\dif+\omega)\wedge e^+$. Signs and factors of the Hodge-$\ast$
operation are defined by $\Hodge\ \epsilon=1$. The auxiliary fields
$X,X^a$ can be interpreted as Lagrange multipliers for geometric
curvature and torsion, respectively. $X^\pm$ correspond to the expansion
spin coefficients $\rho,\rho'$ (both are real in case of spherical symmetry, see below). 

All classical solutions can be obtained with particular ease from (\ref{cs:1}) not only locally, but globally \cite{Klosch:1996fi}. 

Even in the presence of torsion the reduced equations of motion enforce
vanishing torsion unless matter couplings to torsion exist. Such a
discussion will be postponed because fermion fields -- which are the
topic of the next section -- will be needed as sources. It will turn out
that the field $k_a$, the contortion vector,  decouples from the theory 
even in the presence of fermions.

\subsection{Dilaton gravity with matter}
\label{se:3.2}

In this subsection dilaton gravity with matter is discussed. Although
we specialise to spherically reduced gravity the following is still valid
for generic dilaton gravity theories, which means for generic functions $U(X)$
and $V(X)$, combined in the potential $\mathcal{V}=X^+X^-U+V$.

Spherical reduction produces (\ref{cs:1}) with 
$\mathcal{V}=-X^+X^-/(2X)-\la^2$, 
where $\la$ is a physical parameter which can be scaled to 1 by redefining 
the units. By a conformal transformation $e^a\to \tilde{e}^a=e^a \Om$ with 
conformal factor $\Om = X^{1/4}$ the transformed dilaton potential 
$\tilde{V}=-2\la^2\sqrt{X}$ becomes independent of $X^\pm$.
Choosing such a conformal frame is often helpful, however we will not
specify the conformal frame for the time being.

It will be assumed that $X^+\neq 0$ in a given patch. 
If $X^+=0$ and $X^-\neq 0$ everything can be repeated 
with $+\leftrightarrow -$. If both $X^+=0=X^-$ in an open region 
a constant dilaton vacuum is encountered, which will not be discussed 
here (but they are rather trivial anyhow). If $X^+=0=X^-$ at an isolated 
point typically this corresponds to a bifurcation 2-sphere. This slight 
complication will be neglected here as it is not essential for the 
present discussion.\footnote{One can describe a patch in which $X^+=0=X^-$
at a certain point e.g.~by a coordinate system similar to the one
introduced by Israel~\cite{Israel:1966} or by Kruskal like coordinates.}
$X^+X^-=0$ corresponds to an apparent horizon, which in the static case
is a Killing horizon. 

The generic Ansatz for the energy-momentum 1-form is
\begin{equation}
      \label{eq:pf1}
      W^\pm = W^\pm_X \dif X + W^\pm_Z Z\,,
\end{equation}
where the 1-form $Z$ is defined by
\begin{equation}
      \label{eq:pf2}
      Z:=\frac{e^+}{X^+}\,.
\end{equation}
For the following it will make sense to further specify (\ref{eq:pf1}):
\begin{alignat}{2}
      W^+_X &= X^+T_1\,,&\qquad W^+_Z &= X^+T_2\,,
      \nonumber \\
      W^-_X &= X^-T_3\,,&\qquad W^-_Z &= X^-T_4\,,
      \label{eq:pf3}
\end{alignat}
which is only allowed in the absence of horizons.
The EOM
\begin{align}
      & \dif X + X^-e^+ - X^+e^- = 0\,, 
      \label{eq:pf4}\\
      & (\dif\pm\om)X^\pm\mp \mathcal{V}e^\pm + W^\pm = 0\,,
      \label{eq:pf5}\\
      & \dif\om+\frac{\partial\mathcal{V}}{\partial X}\epsilon 
      +W\epsilon = 0\,, 
      \label{eq:pf6}\\
      & (\dif\pm\om)\wedge e^\pm 
      +\frac{\partial\mathcal{V}}{\partial X^\mp}\epsilon= 0\,. 
      \label{eq:pf7}
\end{align}
Let us emphasise that $\omega$ is the Levi-Civit\'{a} connection
only in a conformal frame with $U=0$, i.e.,~$\mathcal{V}=V(X)$.
Together with (\ref{eq:pf1}) and (\ref{eq:pf3}) immediately 
imply the following relations:
\begin{align}
      & e^- = \frac{\dif X}{X^+} + X^-Z\,, 
      \label{eq:pfa}\\
      & \epsilon=e^+\wedge e^-=Z\wedge \dif X\,,
      \label{eq:pfb}\\
      & \om = -\frac{\dif X^+}{X^+} + \mathcal{V}Z - \frac{W^+}{X^+}\,,
      \label{eq:pfc}\\
      & \dif Z = (T_1+U(X)) \dif X\wedge Z\,,
      \label{eq:pfd}
\end{align}
where in addition 
\begin{multline}      
      \dif(X^+X^-)+V(X)\dif X+X^+X^-U(X)\dif X\\
      + X^+X^-(T_1+T_3)\dif X+X^+X^-(T_2+T_4)Z = 0\,,
      \label{eq:pfe}
\end{multline}
indicates the existence of a conserved quantity.
The line element can easily be computed to be
\begin{align}
      \dif s^2 = 2 e^{+}\otimes e^{-} =
      2 X^{+}X^{-} Z \otimes Z + 2 Z \otimes \dif X\,,
      \label{eq:pff}
\end{align}
which follows from~(\ref{eq:pf2}) and~(\ref{eq:pfa}) and 
takes the usual Eddington-Finkelstein gauge,
\begin{align}
      g_{\alpha\beta}=
      \begin{pmatrix}
      2 X^{+}X^{-} & 1 \\
      1 & 0 
      \end{pmatrix}\,,\qquad
      g^{\alpha\beta}=
      \begin{pmatrix}
      0 & 1 \\
      1 & - 2 X^{+}X^{-}
      \end{pmatrix}\,.
\end{align}
By virtue of the previous relations one obtains the 
minus part of~(\ref{eq:pf7})
\begin{multline}
      \dif(X^+X^-)\wedge Z+V(X)\dif X+X^+X^-U(X)\dif X\\
      +(2X^+X^-T_1-T_2) \dif X\wedge Z=0\,,
\end{multline}
which together with~(\ref{eq:pfe}) implies
\begin{align}
      T_2 = X^+X^-(T_1-T_3)\,.
      \label{eq:pf8}
\end{align}
Therefore we are left with three independent functions. This is of course
expected since any symmetric two-dimensional energy-momentum tensor
has only three independent components.

For later use it is important to relate the generic energy-momentum
1-form~(\ref{eq:pf1}) with the energy-momentum tensor $T^{\alpha\beta}$ 
obtained by varying the matter Lagrangian with respect to the metric. 
It would be tempting to vary the matter Lagrangian with respect to the
1-form $e^a$ and relate this object directly with the energy-momentum tensor.
However, in the four dimensional case variation of the matter Lagrangian
with respect to the 1-form $e^m$ gives a 3-form and its dual defines the
energy-momentum tensor. Therefore one finds
\begin{multline}
      \delta L^m = \int \delta e^a \wedge W_{a}
      = \frac{1}{2}\int \delta_{\alpha\beta}^{\gamma\delta}
        \delta  e^a_{\gamma} W_{a\delta}\,\dif x^{\alpha}\wedge \dif x^{\beta}\\
      = \int W_{a\alpha}\eps^{\alpha\beta}\delta e^a_{\beta} \epsilon 
      = \int T_a^{\alpha} \delta e^a_{\alpha} \epsilon \,,
\end{multline}
where $\delta_{\alpha\beta}^{\gamma\delta}$ is the permutation symbol,
and which together with~(\ref{eq:pf8}) implies the following form the 
energy-momentum tensor $T^{\alpha\beta}$ are related by
\begin{align}
      T^{\alpha\beta} = \eps^{\gamma\alpha}W^a_{\gamma}E_a^{\beta}\,.
\end{align}
Hence we find the following energy-momentum tensor
\begin{align}
      T^{\alpha\beta}=
      \begin{pmatrix}
      \ T_1 & -T_2 \\ -T_2 & X^+X^-(T_2-T_4)
      \end{pmatrix}\,.
      \label{eq:pf9}
\end{align}
Note again, that we are working in the Eddington-Finkelstein gauge.

So far all EOMs have been exploited except for two; one of them yields the 
local Lorentz angle (i.e.,\ it determines the ratio of $X^-/X^+$), which 
is not of interest here, while the other one yields the dilaton
current $W$. In addition to the equations of motion one has one more
equation, namely the covariant conservation of the energy-momentum tensor.
In the non-static case we cannot do much more but if in addition
staticity is assumed, we can solve the equations of motion.

\subsubsection{Static and spherically symmetric matter}
In the following staticity is assumed. Then the 
equations of motion simplify considerably and the conservations
equation~(\ref{eq:pfe}) can be integrated. For static
solutions of generic dilaton gravity models 
cf.~e.g.~\cite{Zaslavsky:1999zh,Bronnikov:2003rf,Grumiller:2004wi}.
Staticity implies that $X^+X^-=X^+X^-(X)$ and $T_i=T_i(X)$. 
Putting this into~(\ref{eq:pfe})
immediately leads to
\begin{align}
      T_2+T_4=0\,.
      \label{eq:pf10}
\end{align}
Equation~(\ref{eq:pf6}), which yields the dilaton current, simplifies to 
\begin{align}
      T'_2+T_1(T_2-V+X^+X^-U)+W=0\,,
      \label{eq:pf11}
\end{align}
where the prime means differentiation with respect to the dilaton.
Furthermore the covariant conservation of the energy-momentum 1-form
takes the following form
\begin{align}
      E_a \contra\left(\dif W^a +\eps^a{}_b \om\wedge W^b\right)=
      \bigl(W+X^+X^-U(T_1+T_3)\bigr)\dif X \,, 
      \label{eq:momcons}
\end{align}
where the above relation~(\ref{eq:pf8}) implied the vanishing
of the $Z$ direction and~(\ref{eq:pf10}), (\ref{eq:pf11}) were
used for simplifications. It should be noted that the $4d$
energy-momentum conservation equation is given by~(\ref{eq:momcons}).
Thus,  one concludes that the non-conservation of the $2d$ energy-momentum
tensor is essentially given by the dilaton current $W$.

The conservation equation~(\ref{eq:pfe}) reads 
\begin{align}
      \dif(X^+X^-)+V(X)\dif X+X^+X^-\bigl(U(X)+T_1+T_3\bigr)\dif X=0\,
      \label{cons}
\end{align}
which suggests the definitions
\begin{align}
      I(X):&=\exp{\int^X (U(y)+T_1(y)+T_3(y))\dif y}\,,\nonumber\\
      w(X):&=\int^X I(y)V(y)\dif y\,,
      \label{eq:pf12}
\end{align}
and the total conserved quantity can be integrated to
\begin{equation}
      \label{eq:pf13}
      C = X^+X^- I(X) + w(X) = \rm const.\,,
\end{equation}
which is precisely the form of ordinary dilaton gravity. The difference is, 
of course, that $I$ and $w$ depend on functions present in the 
energy-momentum tensor. Note: If the term $T_1+T_3$ scales with $1/X^+X^-$
one should redefine the potential $V \mapsto V+T_1+T_3$ and 
leave $U$ unchanged. In the absence of horizons $X^+X^- \neq 0$ this
redefinition is well defined. This point does not change the integrability 
feature.

From~(\ref{eq:pfd}) one finds the $Z$ can be written as
\begin{align}
      Z = e^Q \dif u\,,\qquad Q=\int^X (T_1(y)+U(y))\dif y \,,
      \label{eq:pf14}
\end{align}
which in turn gives 
\begin{align}
      \dif X = e^{-Q} \dif r\,.
      \label{eq:pf15}
\end{align}
Therefore the line element simplifies to
\begin{align}
      \dif s^2 = 2\dif u \dif r + K(X) \dif u^2\,,
      \label{eq:pf16}
\end{align}
where the Killing norm is given by
\begin{multline}      
      K(X) = 2e^{2Q}X^+X^-\\=2\exp\Bigl({\int^X(U(y)+T_1(y)-T_3(y))\dif y}\Bigr)
      \bigl(C-w(X)\bigr)\,.
      \label{eq:pf17}
\end{multline}
This is nothing but the most general solution of dilaton gravity (cf.\ eq.\ (3.26) 
of \cite{Grumiller:2002nm}). The aspect that static matter solutions
can be mapped on ordinary solutions of dilaton gravity is discussed 
in subsection~\ref{se:ssspf}, where matter is assumed to be a
static and spherically symmetric perfect fluid.

\subsubsection{Spherically symmetric perfect fluids}
\label{se:sspf}

A perfect fluid is characterised by
\begin{align}
      T^{\mu\nu} = (\rho + P)u^{\mu} u^{\nu} - P g^{\mu\nu}\,,      
      \label{eq:pf18}
\end{align}
where $\rho$ and $P$ denote the energy density and pressure 
respectively with respect to the equal time frame (momentaneous)
defined by $u^{\mu}$, the fluid's four-velocity. The gravitational
field equations imply the vanishing of the covariant derivative
of the energy-momentum tensor
\begin{align}
      \nabla_{\nu} T^{\mu\nu} =0\,.
      \label{eq:pf18a}
\end{align}
Its form is best known in spherically symmetric four-dimensional gravity
in diagonal gauge $\dif s^2=e^{\nu}\dif t^2 -e^a \dif r^2 -r^2\dif\Omega^2$. 
By suppressing the spherical components it reads
\begin{align}
      T^{\mu\nu} = 
      \begin{pmatrix}
      \rho e^{-\nu} & 0 \\ 0 & P e^{-a}
      \end{pmatrix}\,,
      \label{eq:pf19}
\end{align} 
which in the Eddington-Finkelstein gauge becomes
\begin{align}
      T^{\mu\nu} = 
      \begin{pmatrix}
      \frac{\rho + P}{K} & -P \\ -P & P K
      \end{pmatrix}\,.
      \label{eq:pf20}
\end{align} 
Comparing the above energy-momentum tensor with~(\ref{eq:pf9})
leads to the following identifications
\begin{alignat}{2}
      T_1 &= \frac{\rho + P}{2X^+X^-}\,, &\qquad T_2 &= P\,, \nonumber \\
      T_3 &= \frac{\rho - P}{2X^+X^-}\,, &\qquad T_4 &= -P\,.
      \label{eq:pf21}
\end{alignat}
Therefore one immediately finds that $T_1+T_3$ scales with $1/X^+X^-$
which henceforth must be taken into account if the conservation
equation~(\ref{eq:pfe}) is considered, see the discussion above.

Lastly we denote the explicit form of the energy-momentum 
1-form~(\ref{eq:pf3}) for a perfect fluid
\begin{align}
      W^\pm = \mp\frac{\rho-P}{2}e^\pm 
      \pm\frac{X^\pm}{X^\mp}\frac{\rho+P}{2}e^\mp \,,
      \label{eq:pf21a}
\end{align}
which explicitly depends on $X^{\pm}$. It is not surprising that
we cannot recover a perfect fluid action from~(\ref{eq:pf21a}).
However, with prescribed equation of state the action is given by 
the pressure and one can well define what is meant by an 
action principle~\cite{Schutz:1970}.

\subsubsection{Static and spherically symmetric perfect fluids}
\label{se:ssspf}

Spherically symmetric static perfect fluid solutions have been studied in 
several publications, cf.~\cite{Rahman:2001hp} and references therein. 
As may be expected, the discussion becomes particularly easy within the 
reduced theory. The fact that a perfect fluid couples minimally to the dilaton 
also in the reduced theory is a crucial technical ingredient. Assuming 
staticity the EOMs are solved. Integrability of this system can be deduced 
from a general discussion \cite{Mann:1993yv}, but it will be made explicit below.

By virtue of the identification~(\ref{eq:pf12}) one obtains
\begin{align}
      X^+X^-(T_1+T_3)=\rho \,.
      \label{eq:pf22a}
\end{align}
Assume $\rho>0$, then the latter equation~(\ref{eq:pf22a})
implies the absence of Killing horizons, $X^+X^-\neq 0$,
if $|T_1+T_3|<\infty$ holds. However at the boundary of
a perfect fluid sphere the energy density may vanish $\rho=0$.
Since the exterior spacetime is matterless, i.e.,~$T_1+T_3=0$,
there is no horizon located at the boundary. The condition
$X^+X^-\neq 0$ is weaker than the Buchdahl inequality~\cite{Buchdahl:1959}
but suffices to show the non-existence of horizons.

The static conservation equation~(\ref{cons}) gives
\begin{align}
      \dif(X^+X^-) + V(X)\dif X + X^+X^-U(X)\dif X + \rho(X)\dif X =0\,,
      \label{eq:pf22}
\end{align}
where we now see that rather than~(\ref{eq:pf12}) one must choose 
\begin{align}
      I(X):&=\exp{\int^X U(y)\dif y}\,,\nonumber\\
      w(X):&=\int^X I(y)(V(y)+\rho(y))\dif y\,,
      \label{eq:pf23}
\end{align}
which yields the total conserved quantity~(\ref{eq:pf13}) to be
\begin{align}
      C = X^+X^- I(X) + w(X)\,.
      \label{eq:pf25}
\end{align}
The usual energy-momentum conservation~(\ref{eq:pf18a}) is
encoded in equation~(\ref{eq:pf11}) with vanishing dilaton current,
$W=0$. Hence we conclude that a static perfect fluid couples
minimally to dilaton gravity. The conservation equation~(\ref{eq:pf18a})
reads
\begin{align}
      P'+T_1(P-V+X^+X^-U)&=0\,,\nonumber \\
      T_1(P-V+X^+X^-U)&=\frac{K'}{2K}(\rho+P)\,,
      \label{eq:pf27}
\end{align}
where the second relation of~(\ref{eq:pf27}) can be obtained by 
differentiating~(\ref{eq:pf17}) and using the identification~(\ref{eq:pf21}).
The Killing norm for static perfect fluids becomes
\begin{align}
      K(X)=2X^+X^-\exp\Bigl(2\int^X T_1(y)+U(y)\dif y\Bigr)\,,
      \label{eq:pf26}
\end{align}
where we explicitly see that $X^+X^-=0$ corresponds to a Killing
horizon. The identification of $T_1$ can be re-expressed with~(\ref{eq:pf25}) 
and yields
\begin{align}
      T_1 = \frac{\rho+P}{2X^+X^-} = \frac{1}{2}
      \frac{I(X)(\rho+P)}{C-\int^XI(y)(V(y)+\rho(y))\dif y}\,.
      \label{eq:pf28}
\end{align}

It should be noted that the energy-momentum conservation equation 
of an {\it anisotropic} perfect fluid reads
\begin{align}
      P' + \frac{K'}{2K}(\rho+P) = \frac{X'}{X}(P_{\perp}-P)\,,
\end{align}
where $P_{\perp}$ is the orthogonal pressure component of the
anisotropic perfect fluid 
$T^{\mu}_{\nu}=\mbox{diag}(\rho,-P,-P_{\perp},-P_{\perp})$. 
If $P_{\perp}=P$ then the conservation
equations decouples from the dilaton and one is back at the 
isotropic case. Therefore one concludes that an anisotropic
fluid can only be described with non-minimal coupling to the 
dilaton.

Therefore every static, spherically symmetric, minimally coupled ($W=0$) matter 
solution can be mapped onto solutions of a dilaton gravity model, 
see the discussion that follows. This in particular includes the discussed perfect
fluid case. For a colliding null dust this statement can already be 
found in~\cite{Gergely:1998ba}, in our framework this corresponds
to prescribing the pressure to vanish.

Starting from~(\ref{eq:pf16}) and~(\ref{eq:pf17}) with the redefinition
\begin{align}
      \dif r = \exp\Bigl(\int^{\tilde{X}}(U(y)+T_1(y)-T_3(y))\dif y\Bigr)\dif\tilde{X}
      =\tilde{I}(\tilde{X})\dif\tilde{X}\,,
\end{align}
yields the line element in the following form
\begin{align}
      \dif s^2 = \tilde{I}(\tilde{X})\bigl(
      2\dif u \dif\tilde{X} +(C-\tilde{w}(\tilde{X}))\dif u^2\bigr)\,,
\end{align}
where we furthermore redefined~(\ref{eq:pf12}) to be
\begin{align}
      \tilde{w}(\tilde{X})&=\int^{\tilde{X}}\tilde{I}(y)\tilde{V}(y)\dif y\,,
      \nonumber\\
      \tilde{V}(\tilde{X})&=V(\tilde{X})
      \exp\Bigl(2\int^{\tilde{X}}T_3(y)\dif y\Bigr)\,.
\end{align}
Let us now, in contrast to the perfect fluid case, assume
that the function $T_1(X)$ is given, which corresponds
to the introduction of some generating function~\cite{Rahman:2001hp}.
Note that for given $T_1(X)$ equation~(\ref{eq:pf11}) yields $T_2(X)$ and
therefore $T_4(X)$ by~(\ref{eq:pf10}) and finally $T_3(X)$ is 
obtained from~(\ref{eq:pf8}).

Hence for each choice of $T_1$ in the dilaton gravity sector there is 
exactly one $\tilde{w}$ in the matter or perfect fluid sector. 
However, not every $\tilde{w}$ permits a {\em regular} representation 
as a perfect fluid! Only if one allows for singular configurations 
all\footnote{By ``all'' in the parlance of \cite{Grumiller:2003dh} it is meant that 
the ``good'' function $\tilde{w}$ can attain any form. The ``muggy'' function 
$\tilde{I}$, however, cannot be chosen independently.} $2d$ dilaton gravity theories 
can be mapped onto a static spherically symmetric perfect fluid model coupled to 
Einstein gravity in $d=4$. This can be seen most easily be checking that for 
regular $T_1$ the relation between $X$ and $\tilde{X}$ is invertible. The same 
holds for $r$ and $X$. Thus, these three coordinates can be expressed as 
monotonous functions with respect to each other (e.g.\ $X(\tilde{X})$). 
Because the original $V(X)$ is also monotonous, this means that also $\tilde{w}$ 
is monotonous. Moreover, the function $\tilde{I}$ cannot be zero. Therefore, 
there can be at most one (non-extremal) Killing horizon, depending on the sign 
of $C$ and eventual lower or upper bounds of $\tilde{w}$. Thus, the only 
possibility to express generic $2d$ dilaton gravity as a perfect fluid model 
is to allow for singular energy distributions. However, at a singular point of 
$T_1$ all previous coordinate redefinitions are not valid anymore. Only if one 
simultaneously performs a conformal transformation with compensating 
singularities---thus changing the causal structure in an essential way---
finally all dilaton gravity models can be reproduced.

In this sense, generic $2d$ dilaton gravity corresponds to a 
(not necessarily regular) perfect fluid solution in a certain 
(not necessarily regular) conformal frame. However, regardless of this 
minor interpretational issue the particular ease of this formalism 
should be emphasised and compared with the usually more involved 
calculations in $d=4$.

\subsubsection*{Further remarks and comments}
In order to complete the perfect fluid discussion some remarks are
necessary. Firstly one should have in mind that the Einstein
field equations for a static and spherically symmetric perfect fluid
reduce to a system of two first order differential equations for 
a given equation of state. Existence and uniqueness of the solution
of this system was proved in~\cite{Rendall:1991hg} for an already wide class
of equations of state. Many assumptions on the equation of state could
later be weakened in~\cite{Baumgarte:1993} and~\cite{Mars:1996gd}.

The power of dilaton gravity is to get equations of motion of the
first order, so it seems that in the perfect fluid case only little
can be won, namely the total conserved quantity $C$ in~(\ref{eq:pf25}).
The disadvantage on the other hand is the more complicated structure
of the differential equations if the density, the pressure or an equation
of state is specified. Already in the constant density case equations
get more involved than with the usual approach through the 
Tolman-Oppenheimer-Volkoff~\cite{Tolman:1939jz,Oppenheimer:1939ne}
equation. As expected, the three equations~(\ref{eq:pf25})--(\ref{eq:pf27}) 
contain four unknown functions, namely $\rho, P, K$ and $X^+X^-$, 
therefore one of these functions can be chosen freely. 

\subsection{Spinor formalism and reduction of fermions}
\label{subsec_red_fer}

Since spherical symmetry provides a foliation of spacetime by spacelike
two-surfaces (round two-spheres) it is natural to adapt the Clifford
algebra to this foliation. In particular the Geroch-Held-Penrose (GHP)
spin-coefficient formalism~\cite{Geroch:1973,penrosespinors} 
is particularly well suited for this situation.
It uses a double-null tetrad $(l^{a},n^{a},m^{a},\bar{m}^{a})$ 
satisfying\footnote{We note that we follow here the usage of most
of the literature, taking Latin indices for abstract indices, whereas
they were used for anholonomic indices in the previous section.}
\begin{align}
      &l\cdot n=1\,,\,m\cdot\bar{m}=-1\,,\qquad 
      l^{2}=0\,,\, n^{2}=0\,,\,m^{2}=0\,,\bar{\, m}^{2}=0\,,\\
      &m_a \dif x^a=-\frac{\Phi}{\sqrt{2}}(\dif\theta-i\sin\negthinspace\theta\dif\phi)\,,
\end{align}
adapted to such a foliation by noticing that the orthogonal complement
of the tangent space of the two-surfaces is uniquely spanned by two
null-normals $l^{a},\, n^{a}$. 
\\
\begin{figure}[ht!]
\begin{center}
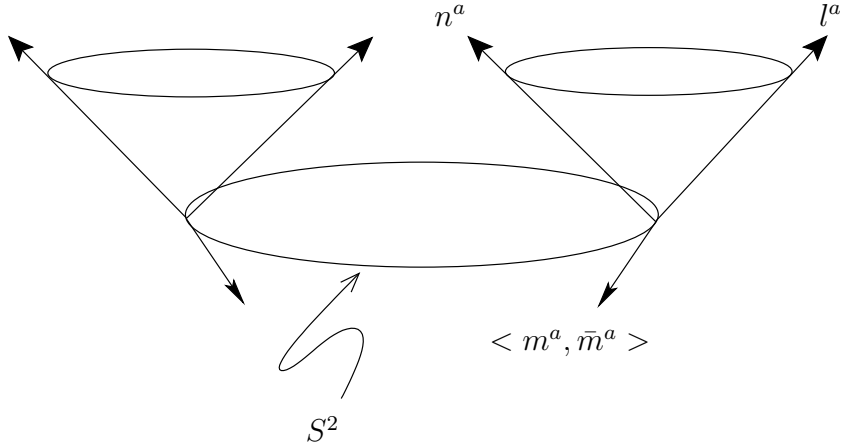
\end{center}
\caption{Foliation of spacetime by two-spheres}
\label{fig:2dspin}
\end{figure}

In the GHP formalism the null tetrad
gives uniquely rise to a spinor basis (dyad) via the identification
\begin{align}
      &l^{a}=o^{A}o^{A'}\,,\, n^{a}=\iota^{A}\iota^{A'}\,,\,
      m^{a}=o^{A}\iota^{A'}\,,\,\bar{m}^{a}=\iota^{A}o^{A'}\,,
      \label{nulltetrad}\\
      &g_{ab}=\e_{AB}\e_{A'B'}=l_a n_b+n_a l_b-m_a \bar{m}_b-\bar{m}_am_b\,,
      \label{null_metric}\\
      &\e^{AB}o_{A}\iota_{B}=o_{A}\iota^{A}=1\,.
\end{align}
This identification allows us to consider tensor fields as a special
case of spinor-fields, by identifying a tensor index $a$ with a pair
of primed and unprimed spinor indices $AA'$. Note that in the previous
section Latin letters were used for anholonomic indices, in this section
they are used as abstract indices.

The covariant derivatives along the null directions of the tetrad define 
the 12 complex spin coefficients (taking into account priming and
complex conjugation)
\begin{alignat}{2}
        D o^A &= -\gamma' o^A - \kappa \iota^A\,,&\qquad
        D \iota^A &= \gamma' \iota^A - \tau' o^A\,, \label{defspin2} \\
        \delta o^A &= \beta o^A - \sigma \iota^A\,,&\qquad    
        \delta \iota^A &= -\beta \iota^A - \rho' o^A\,,  \label{defspin4}
\end{alignat}
where $D=l^a\nabla_a$ and $\delta=m^a\nabla_a$. The GHP formalism and
Cartan's form calculus can be linked by noting appendix~\ref{app_link}
and especially equations~(\ref{cartan1}) and~(\ref{cartan2}). They
can be used to read of the spin coefficients for a given null tetrad.

In a spherically symmetric spacetime 6 of the 12 spin coefficients
vanish, see~\cite{Gubser:1997cm} for the static and spherically 
symmetric case. The vanishing coefficients 
are $\kappa = \sigma = \tau = 0$, together with their 
primed counterparts.
Furthermore $\gamma$ and $\gamma'$ are real quantities 
describing the $2d$ spacetime only. $\rho$ and $\rho'$,
which are also real, describe the expansion of the
sphere. As already said in the end of subsection~\ref{subsub_hilbert},
they correspond to $X^{\pm}$ respectively. The remaining 
two spin coefficients are not independent and are explicitly given by 
$\beta=\bar{\beta}'=(\cot\negthinspace\theta)/(2 \sqrt{2} \Phi)$.

The two-spinor equivalent Dirac action functional~\cite{Godina:1999ti} 
can be written
\begin{align}
      L = \int \bigl(i\,\bs{\bar{\Psi}}\rlap{\,/}\nabla \bs{\Psi} 
      -m\bs{\bar{\Psi}}\bs{\Psi}\bigr)\omega_{G}\,,
      \label{d_action}
\end{align}
where $\bs{\bar{\Psi}}=\bs{\Psi}^{\dagger}\gamma^{0}$ is
the Dirac conjugate and $\rlap{\,/}\nabla=\gamma^a\nabla_a$. 
We take the space of Dirac 4-spinors
$\bs{\Psi}$ to be of the form $\bs{\Psi}=(\psi_{A},\chi^{A'})$,
i.e., the direct sum of the dual with the complex conjugate 
2-spinor space. Note that $\psi_{A}$ and $\chi^{A'}$ are left-
and right-handed fermions respectively. 
In this combination the Dirac conjugate 
is simply given by $\bs{\bar{\Psi}}=(\bar{\chi}^{A},\bar{\psi}_{A'})$.
The Clifford algebra associated with $g_{ab}$ is
follows from the identification
\begin{align}
        \rlap{/}v = \sqrt{2} 
        \begin{pmatrix} 0 & v_{AB'} \\ v^{A'B} & 0\end{pmatrix}\,,\qquad
        \rlap{/}u \rlap{/}v + \rlap{/}v \rlap{/}u = 2(u \cdot v)\unity\,.
        \label{vslash}
\end{align}
Identifying the vectors of the null tetrad~(\ref{nulltetrad})
leads to 
\begin{gather}
        \lsl \nsl + \nsl \lsl = 2 \unity\,,\qquad      
        \msl \mbsl + \mbsl \msl = -2 \unity\,.
        \label{clifford}
\end{gather}
Thus the four-dimensional Clifford algebra is generated
by two two-dimen\-sional Clifford algebras. The first
is generated by the orthonormal basis $l^a$ and $n^a $, 
the second by the basis vectors of the 
two-sphere $\sph$, $m^a$ and $\bar{m}^a$. 

Furthermore we find
\begin{align} 
      S_{\I} = \left\{\begin{pmatrix}o_A\\0\end{pmatrix}, 
      \begin{pmatrix}0\\ \iota^{A'}\end{pmatrix}\right\}\,,\qquad
      S_{\II} = \left\{\begin{pmatrix}0\\o^{A'}\end{pmatrix}, 
      \begin{pmatrix}\iota_{A}\\0\end{pmatrix}\right\}\,,
\end{align}
as invariant subspaces of the two-dimensional Clifford algebra
generated by $l^a$ and $n^a$, hence one can write 
$S=S_{\I}\oplus S_{\II}$. With respect to this basis the
two-dimensional Clifford algebra is represented by
\begin{align}
      \lsl\rightarrow\gamma^{-}_{\III}=\pm\sqrt{2}\begin{pmatrix}0&1\\0&0\end{pmatrix}\,,
      \qquad
      \nsl\rightarrow\gamma^{+}_{\III}=\pm\sqrt{2}\begin{pmatrix}0&0\\1&0\end{pmatrix}\,,
\end{align}
from which the $\gamma^a$-matrices in a local Lorentz frame
are given by
\begin{align}
        \gamma^0_{\III}=\pm\begin{pmatrix} 0 & 1 \\ 1 & 0 \end{pmatrix}\,,\qquad
        \gamma^1_{\III}=\mp\begin{pmatrix} 0 & 1 \\-1 & 0 \end{pmatrix}\,, 
        \label{gamma+-} \\
        \gamma^{\star} =\gamma^0 \gamma^1= 
        \begin{pmatrix} 1 & 0 \\ 0 & -1 \end{pmatrix}\,,
        \label{gamma*}
\end{align}
where we used $\sqrt{2}\gamma^0=\gamma^{+}+\gamma^{-}$. The
upper and lower signs refer to the invariant subspaces
$S_{\I}$ and $S_{\II}$ respectively.

From the above the Dirac action functional~(\ref{d_action}) can be written
in two-spinor form
\begin{align}
        L = \int \Bigl( i\sqrt{2}(\bar{\psi}_{A'}\nabla^{AA'}\psi_A+ 
        \bar{\chi}^A \nabla_{AA'} \chi^{A'})-
        m\bar{\psi}_{A'}\chi^{A'}-m\bar{\chi}^{A}\psi_A \Bigr)\omega_{G}\,,
        \label{action}
\end{align}
which by variation with respect to the spinors $\bar{\psi}_{A'}$ and $\bar{\chi}^A$
leads to the Dirac equation~\cite{Dirac:1928} 
in two-spinor form
\begin{align}
        i\sqrt{2}\,\nabla^{AA'}\psi_{A}-m\chi^{A'}=0\,,\qquad 
        i\sqrt{2}\,\nabla_{AA'}\chi^{A'}-m\psi_{A}=0\,.
        \label{eom}
\end{align}
The Dirac two-spinors are expanded in terms of the basis spinors
$\psi^A = A o^A + P \iota^A$, $\chi^{A'} = B \iota^{A'} + Q o^{A'}$,
where $A$, $B$, $P$ and $Q$ are functions of all four 
spacetime coordinates.
The functions $A$, $B$, $P$ and $Q$ have spin weights $-1/2$, $-1/2$,
$1/2$ and $1/2$ respectively. Therefore one can rewrite the first term 
of the Dirac action~(\ref{action}) in terms of weighted derivative 
operators~\cite{penrosespinors}
\begin{align}
      \bar{\psi}_{A'} \nabla^{AA'} \psi_A = 
      \bar{A}(\thorn -\rho)A +
      \bar{A}\eth' P+\bar{P}(\thorn' -\rho')P+ 
      \bar{P}\eth A\,,
\end{align}
where the weighted operators are given by
\begin{align}
      \thorn\eta &= D\eta + \frac{w}{2}(\gamma' +\bar{\gamma}')\eta
                      + \frac{s}{2}(\gamma' -\bar{\gamma}')\eta\,,
      \label{thorn}\\
      \eth \eta &= \delta\eta - \frac{w}{2}(\beta - \bar{\beta}')\eta
                          - \frac{s}{2}(\beta + \bar{\beta}')\eta\,,  
      \label{eth}
\end{align}
when acting on a weighed quantity $\eta$ with spin weight $s$
and boost weight $w$.
It was taken into account that the spin coefficients
$\kappa$, $\sigma$ and $\tau$ together with their primed
counterparts vanish in case of spherical symmetry. 

Since the action~(\ref{action}) must be a real functional the real 
spin coefficients $\rho$ and $\rho'$ drop out because of 
the factor $i$ in front. This yields
\begin{multline}
        L = \int \Bigl( i\sqrt{2} 
        \bigl(\bar{A}\thorn A + \bar{A}\eth' P+ 
        \bar{B}\thorn' B + \bar{B} \eth' Q\bigr)\\
        i\sqrt{2}\bigl(
        \bar{Q}\thorn Q + \bar{Q}\eth B +
        \bar{P} \thorn' P + \bar{P} \eth A
        \bigr)\\
        - m\bigl(\bar{A}B +\bar{B}A\bigr)
        + m\bigl(\bar{Q}P +\bar{P}Q\bigr) \Bigr)
        \omega_{g} \Phi^2 \dif^2 \Omega\,.
        \label{action2}
\end{multline}
Next the weighed functions $A$, $B$, $P$ and $Q$ are expanded 
in terms of spin weighted spherical harmonics with the appropriate spin 
weights, $A=\sum_{jm}A_{jm}{}_{-\frac{1}{2}}Y_{\frac{1}{2}\,\frac{1}{2}}$, etc. 
These are the eigenfunctions of the operator $\eth' \eth$ for each spin weight $s$. 
They are defined by~\cite{Goldberg:1967uu,penrosespinors}
\begin{align}
        \eth' \eth _s Y_{j,m} &= -\frac{(j+s+1)(j-s)}{2 \Phi^2} {}_s Y_{j,m}\,,  
        \label{eigeny} \\
        \eth _s Y_{j,m} &= -\frac{\sqrt{(j+s+1)(j-s)}}{\sqrt{2} \Phi} {}_{s+1} Y_{j,m}\,,
        \label{ethy} \\
        \eth' _s Y_{j,m} &= \frac{\sqrt{(j-s+1)(j+s)}}{\sqrt{2} \Phi} {}_{s-1} Y_{j,m}\,,
        \label{eth'y}
\end{align} 
and, for each spin weight $s$, enjoy the orthogonality condition
\begin{align}
      \left \langle {}_s Y_{j,m} ,\ {}_s Y_{j',m'} \right \rangle 
      = \frac{1}{4 \pi} \delta_{jj'} \delta_{mm'}\,,\qquad
      \left \langle f,g \right \rangle = 
      \frac{1}{4 \pi}\int\bar{f}g\dif^2 \Omega\,.
      \label{ortho}
\end{align}

Hence the spherical dependence of (\ref{action2}) can be integrated out. 
In particular we obtain
\begin{align}
        \int \bar{A} \thorn A \dif^2 \Omega
        &=\sum_{jm} \bar{A}_{jm} \thorn A_{jm}\,,
        \label{int1} \\
        \int \bar{A} \eth' P \dif^2 \Omega
        &=\frac{j+\frac{1}{2}}{\sqrt{2}\Phi}\sum_{jm}\bar{A}_{jm} P_{jm}\,.
        \label{int2}
\end{align}
Then the spherically reduced fermion action reads
\begin{multline}
        L_{\rm D}=\sqrt{2}\sum_{jm} \int \Bigl(
        i\bigl(\bar{A}_{jm}\thorn A_{jm}
        +\frac{j+\frac{1}{2}}{\sqrt{2}\Phi}\bar{A}_{jm} P_{jm}\bigr)+ 
        i\bigl(\bar{B}_{jm} \thorn' B_{jm} 
        +\frac{j+\frac{1}{2}}{\sqrt{2}\Phi}\bar{B}_{jm} Q_{jm}\bigr)\\+
        i\bigl(\bar{Q}_{jm} \thorn Q_{jm} 
        -\frac{j+\frac{1}{2}}{\sqrt{2}\Phi}\bar{Q}_{jm} B_{jm}\bigr)+
        i\bigl(\bar{P}_{jm} \thorn' P_{jm} 
        -\frac{j+\frac{1}{2}}{\sqrt{2}\Phi}\bar{P}_{jm} A_{jm}\bigl)\\
        -\frac{m}{\sqrt{2}}\bigl(\bar{A}_{jm}B_{jm}+\bar{B}_{jm}A_{jm}\bigr)
        +\frac{m}{\sqrt{2}}\bigl(\bar{P}_{jm}Q_{jm}+\bar{Q}_{jm}P_{jm}\bigr)
        \Bigr) \Phi^2 \omega_{g}\,.
        \label{actionred}
\end{multline}

\subsection*{Two-spinor representation}

Let the two-spinors with respect to their invariant subspace $S_{\III}$ 
respectively be\footnote{If the spinor $\psi_A$ is
labelled by spinorial indices~$A, A',\ldots$ 
then $\bar{\psi}_{A'}$ denotes the complex conjugate of the
spinor $\psi_A$. In all other cases the quantity $\bar{\Psi}$ 
is the Dirac conjugate of the spinor $\Psi$. } 
\begin{align}
        \Psi^{\I}_{jm} &= \begin{pmatrix} A_{jm} \\ B_{jm} \end{pmatrix}\,,\qquad
        \bar{\Psi}^{\I}_{jm} = \Psi^{\I\dagger}_{jm} \gamma^{0}_{\I} = 
        \begin{pmatrix} \bar{B}_{jm}\, ,& \bar{A}_{jm} \end{pmatrix}\,,
        \label{psiplus} \\
        \Psi^{\II}_{jm} &= \begin{pmatrix} Q_{jm} \\ P_{jm} \end{pmatrix}\,,\qquad
        \bar{\Psi}^{\II}_{jm} = \Psi^{\II\dagger}_{jm} \gamma^{0}_{\II} = 
        -\begin{pmatrix} \bar{P}_{jm}\, ,& \bar{Q}_{jm} \end{pmatrix}\,.
        \label{psiminus}
\end{align}

The weighted derivative operators $\thorn$ and $\thorn'$ are simply given
by $\thorn = l^a \nabla_a$ and $\thorn' = n^a \nabla_a$. 
Therefore the dyads $E_{+}^a = l^a$ and $E_{-}^a = n^a$
in light cone form are introduced.

Thus the reduced action~(\ref{actionred}) written in an 
intrinsically $2d$ form becomes
\begin{multline}
        L_{\rm D} = \sum_{jm} \int \Bigl( 
        \bar{\Psi}^{\I}_{jm} \bigl(
        i E_{+}^a\nabla_a\gamma_{\I}^{+}+i E_{-}^a\nabla_a\gamma_{\I}^{-}-m\unity 
        \bigr) \Psi^{\I}_{jm}+\\
        \bar{\Psi}^{\II}_{jm} \bigl(
        i E_{+}^a\nabla_a\gamma_{\II}^{+}+i E_{-}^a\nabla_a\gamma_{\II}^{-}-m\unity 
        \bigr) \Psi^{\II}_{jm}+\\ 
        \frac{j+\frac{1}{2}}{\Phi}\bigl(
        \bar{\Psi}^{\II}_{jm}\gamma^{\star}I\Psi^{\I}_{jm}+\bar{\Psi}^{\I}_{jm}
        \gamma^{\star}I^{-1}\Psi^{\II}_{jm}\bigr)\Bigr)\Phi^2 \omega_{g}\,,
        \label{actionred2}
\end{multline}
where $I$ is the intertwiner between the representations of the two-dimensional
Clifford algebra in $S_{\I}$ and $S_{\II}$ respectively.
The unity matrices in the first and second line should also 
carry an index with respect to their spinor-space, which was avoided 
for clarity. With respect to the bases chosen in $S_{\I}$ 
and $S_{\II}$ we have
\begin{align}
      I:S_{\I}\rightarrow S_{\II}\,,\qquad 
      I=\begin{pmatrix} i & 0\\0 & -i \end{pmatrix}=i\gamma^{\star}\,.
\end{align}
This allows us to identify $S_{\I}$ with the two-dimensional
(irreducible) representation space $S$ of the (two-dimensional)
Clifford algebra, whereas the representation in $S_{\II}$ is
equivalent under the action of the intertwiner $I$. We rewrite
the reduced action by denoting $\Psi^{\I}_{jm}=\eta$ (this means
identifying $S$ with $S_{\I}$) and $\Psi^{\II}_{jm}=I\lambda$ where
$\lambda \in S$. Then the above expression~(\ref{actionred2})
turns into
\begin{align}
      L = \int \Bigl(
      \bar{\eta}\bigl(i\rlap{\,/}\nabla -m\unity 
      \bigr) \eta+
      \bar{\lambda}\bigl(i\rlap{\,/}\nabla -m\unity 
      \bigr) \lambda+   
      \frac{j+\frac{1}{2}}{\Phi}\bigl(
      \bar{\lambda}\gamma^{\star}\eta-\bar{\eta}\gamma^{\star}\lambda
      \bigr)\Bigr)\Phi^2 \omega_{g}\,,
\end{align}
where the summation over the modes is understood and henceforth the
kinetic term is abbreviated using 
$\rlap{\,/}\nabla =E_{+}^a\nabla_a\gamma_{\III}^{+}+E_{-}^a\nabla_a\gamma_{\III}^{-}$
when acting on $\eta$ or $\lambda$, respectively. In this formulation 
both spinors $\eta$ and
$\lambda$ belong to the same spinor space. Finally we
introduce an internal index $u$ and write $\psi_u=(\eta,\lambda)$
\begin{align}
      L = \int \Bigl(\delta^{uv}\bar{\psi}_u\bigl(i\rlap{\,/}\nabla -m\unity 
      \bigr)\psi_{v}-\frac{j+\frac{1}{2}}{\Phi}\e^{uv}\bigl(
      \bar{\psi}_u\gamma^{\star}\psi_{v}\bigr)\Bigr)\Phi^2 \omega_{g}\,,
\end{align}
which display that the action is in a $SO(2)\simeq U(1)$ covariant form.

This remaining freedom is the two parameter subgroup of the Lorentz
group at each point and can be understood from the following. One
can rescale the basis spinors by a complex scalar field 
$o^A \mapsto\Lambda o^A$ and $\iota^A \mapsto (1/\Lambda)\iota^A$, which leaves
the null directions invariant. By writing $\Lambda^2 = R\exp(i\phi)$ one
finds that the null directions $l^a$ and $n^a$ are boosted, whereas
$m^a$ and $\bar{m}^a$ are rotated by an angle $\pm\phi$, respectively.

\begin{figure}[ht!]
\begin{center}
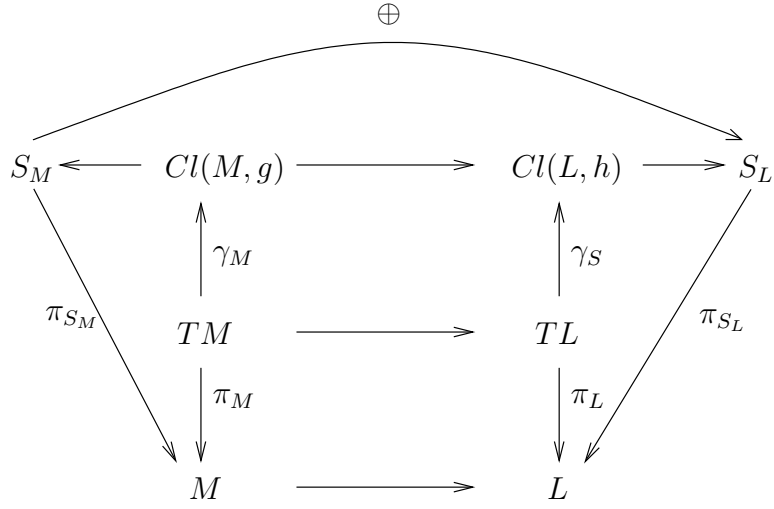
\end{center}
\caption[Embedding diagram for the spin-spaces]{Embedding diagram for the spin-spaces of 
$\mathcal{M}={M,L}$. $T{\mathcal{M}}$ denotes
the tangent bundle and $Cl(\mathcal{M},g_\mathcal{M})$ the corresponding
Clifford bundle with Clifford-map $\gamma_\mathcal{M}$. 
$S_\mathcal{M}$ refers to the respective spin-bundles, 
i.e., representation spaces of $Cl(\mathcal{M},g_\mathcal{M})$.}
\label{fig:2dspin_cd}
\end{figure}

Graphically the above spaces and their embeddings
are summarised in figure~\ref{fig:2dspin_cd}.

\subsection{Reduction of Yang-Mills fields}

In this subsection we use the metric formalism to spherically reduce
Yang-Mills fields, where Latin letters correspond to abstract indices.
This formalism uses a generic metric and does not specify the
signature, in contrast to the GHP formalism, in which the signature
is fixed.

Before starting with the nonabelian case it is worthwhile to consider $U(1)$. In standard notation \cite{penrosespinors} the skew-symmetric field tensor
\begin{align}
      F_{ab} = \phi_{AB}\eps_{A'B'}+\eps_{AB}\bar{\phi}_{A'B'}\,,
      \label{eym1}
\end{align}
can be decomposed in terms of a complex, symmetric bispinor $\phi_{AB}$. If a potential exists the relation $\phi_{AB}=\nabla_{A'(A}A_{B)}{}^{A'}$ implies $F_{ab}=2\nabla_{[a}A_{b]}$.

According to our notion, spherical symmetry means that the Lie-derivative taken into the Killing-directions acting on the (in the present case abelian) Yang-Mills action yields only surface terms. A sufficient, but by no means necessary condition is strict spherical symmetry, $\killing_{\xi}F_{ab}\stackrel{!}{=}0$, implying $\killing_{\xi}\phi_{AB}=0$ (since $\killing_{\xi}\eps_{AB}=0$ by construction). More explicitly this condition reads
\begin{align}
      (\xi\cdot\nabla)\phi_{AB} + \Phi_A{}^C\phi_{CB} + \Phi_B{}^C \phi_{AC} = 0\,.
      \label{eym2}
\end{align} 
Applying the decomposition with respect to the basis,
$\phi_{AB}=\phi^{00}o_Ao_B+\phi^{01}o_{(A}\iota_{B)}+\phi^{11}\iota_A\iota_B$,
yields three conditions for the three coefficients $\phi^{ij}$. In the
same way in which the nonexistence of strictly spherically symmetric spinors can be proved the relation $\phi^{ii}=0$ can be shown. However, as opposed to spinors this does not imply a trivial field configuration. Indeed, the equation
\begin{align}
      (\xi\cdot\nabla)(\phi^{01}o_{(A}\iota_{B)})=0\,,
      \label{eym3}
\end{align}
after contraction with $o^B\iota^A$ reduces to
\begin{align}
      (\xi\cdot\nabla)\phi^{01} = 0\,,
      \label{eym4}
\end{align}
allowing for nontrivial field configurations (namely electric monopoles and their dual). This is of course not unexpected, since the Coulomb-solution is well-known for exhibiting spherical symmetry. This result can be generalised to the Yang-Mills case. However, the condition of strictness can be relaxed.

Since $su(2)$ is the building block of all other Lie-Algebras (and for
sake of simplicity) we will restrict ourselves to the gauge group
$SU(2)$. It can be expected that spherical reduction yields a
non-trivial result due to the fact that the $SU(2)\equiv S^3$ allows for
a Hopf-fibration $U(1)$ over $S^2$ and because the isometry group of the
metric by construction contains $SO(3)$ as subgroup with
$S^2$-orbits. Spherical reduction of $SU(2)$-Yang-Mills theory has been performed by several authors during the 1970's -- the most prominent is probably reference~\cite{Witten:1977ck}. We will follow in our description closely the approach of Forg\'{a}cs and Manton \cite{Forgacs:1980zs}. The condition
\begin{align}
      \killing_{\xi_m} A_a \stackrel{!}{=} D_a W_m\,,
      \label{eym5}
\end{align}
provides spherical symmetry up to gauge transformations. For each Killing vector $\xi_m$ a Lie-algebra valued scalar field $W_m=W_m^i T^i$ is introduced, with $T^i$ being the generators of $SU(2)$. Note that $m$ is not a usual abstract index, but just a label for the $m^{\rm th}$ Killing vector. $D_a$ is the gauge-covariant derivative, i.e., $D_a W = \nabla_a W -ig[A_a,W]$. Equation.\ (\ref{eym5}) is equivalent to gauge-covariant transformation behaviour of the nonabelian field tensor $\killing_{\xi}F_{ab}=ig[F_{ab},W]$. 

Applying the commutator of two Lie-derivatives establishes the (Wess-Zumino) consistency condition
\begin{align}
      2\killing_{\xi_[m}W_{n]} = [W_m,W_n]+f_{mnl}W_l\,,
      \label{eym6}
\end{align}
with the structure constants $[\xi_m,\xi_n]=if_{mnl}\xi_l$

The general idea to solve (\ref{eym5}) as advocated in~\cite{Forgacs:1980zs} is as follows: instead of solving this equation on the coset-space $S^2=SO(3)/SO(2)$, it is solved using the whole symmetry group $SO(3)$, thus introducing an additional dimension. By a gauge transformation one can simplify (\ref{eym5}) such that the right hand side vanishes. Equations of this type can be solved easily (they must be fulfilled separately for each generator). Afterwards a gauge transformation is used to effectively project the gauge-field to the coset space. Then one proceeds to find the most general solution to the consistency conditions.

What has been explained in words will now be presented briefly in formulas. The abstract index-set $\{\tilde{a},i,w\}$ is split into the isometry part $\tilde{a}$ (``$r-t$-part''), the coset part $i$ (``$\theta-\phi$''-part) and the phase part $w$ (the third Euler angle $\chi=x^w$). A generic index of the subset $\{i,w\}$ is denoted with $\hat{a}$. A generic index of the subset $\{\tilde{a},i\}$ is denoted by $a$. The solution of
\begin{align}
      \killing_{\xi_m} A_{\hat{a}}^i = 0\,,\qquad\forall m,i\,, 
      \label{eym7}
\end{align}
reads
\begin{align}
      A_{\hat{a}}^i=\Phi^i_m\tilde{\xi}_{m\hat{a}}\,,
      \label{eym8}
\end{align}
where $\tilde{\xi}$ denotes left-translations (as opposed to the usual right-translations $\xi$). The scalars $\phi^i_m$ are independent of the phase $x^w$. The $\tilde{a}$-component is trivially given by
\begin{align}
      A_{\tilde{a}}^i=A_{\tilde{a}}^i(x^{\tilde{a}})\,.
      \label{eym9}
\end{align}
Afterwards a gauge is chosen such that $A_w=0$ and the other components be independent of $x^w$ (so in fact the third Euler angle decouples and one can restrict to the coset space).

For explicit calculations one assumes that the generator of the $SO(2)$ subgroup (defining the coset space $S^2$) is $T^3$. There are three possible solutions of the consistency equations
\begin{align}
      \nabla_a \Phi^i_3 -g\eps^{i\be\ga}A_a^\be\Phi_3^\ga&=0\,, \\
      \eps_{m3l}\Phi^i_l+g\eps^{i\be\ga}\Phi^\be_m\Phi^\ga_3&=0\,. 
      \label{eym10}
\end{align}
It is convenient to fix a gauge where $\Phi^1_3=\Phi^2_3=0$. 

There exists a degenerate solution when all $\Phi^i_m$ vanish, which produces simply pure $SU(2)$ gauge theory after reduction (this is the analogue of the $U(1)$-example discussed before). Another solution is obtained for vanishing $\Phi_1=\Phi_2$ and constant $\Phi_3^3$, corresponding to an abelian monopole of arbitrary charge.

The solution of the constraint equations in the general case is given by
\begin{align}
      A_a^i &=\frac{1}{g}(0, 0, a_a)\,,\\
      \Phi_1^i &=\frac{1}{g} (\Phi_1, \Phi_2, 0)\,,\\
      \Phi_2^i &=\frac{1}{g} (\Phi_2, -\Phi_1, 0)\,,\\
      \Phi_3^i &=\frac{1}{g} (0, 0, 1)\,.
\end{align}
A gauge rotation which makes $A^i_w$ equal to zero for all $i$
establishes an equivalent form, the so-called Witten ansatz:
\begin{align}
      A_t^i&=\frac{1}{g}(0,0,a_0)\,,\\
      A_r^i&=\frac{1}{g}(0,0,a_1)\,,\\
      A_\theta^i&=\frac{1}{g}(-\Phi_1,-\Phi_2,0)\,,\\
      A_\phi^i&=\frac{1}{g}(\Phi_2\sin\theta,-\Phi_1\sin\theta,\cos\negthinspace\theta)\,.
\end{align}
It leads after reduction to an abelian gauge theory, supplemented by a
complex scalar field. The corresponding scalar fields $W_m^i$ 
read (cf. equation~(\ref{eym5}))
\begin{align}
      W_1^3 = \frac{\sin\phi}{\sin\negthinspace\theta}\,,\quad 
      W_2^3=\frac{\cos\negthinspace\phi}{\sin\theta}\,,\quad 
      W_m^i = 0 \,\, {\rm otherwise.}
      \label{eym11}
\end{align}

The real scalars $\Phi_1$ and $\Phi_2$ are combined to one complex
quantity $w$. Most conveniently~\cite{Volkov:1998cc}, the spherically 
symmetric Lie algebra valued 1-form can be written as follows:
Let us denote $a=a_0\dif t+a_1\dif r$ and for $SU(2)$ we have 
$\textbf{T}_i=\sigma_i/2$, where $\sigma_i$ are the Pauli matrices.
For $SU(3)$ the above Witten ansatz has to be supplemented 
by just one additional term $(b/2g)\lambda_8$, where 
$b=b_0\dif t+b_1\dif r$, $\textbf{T}_i=(\lambda_1,\lambda_2,\lambda_3)/2$ and 
$\lambda_i$ are the standard Gell-Mann matrices. Therefore
\begin{multline}
      A=\textbf{T}_i A^{i}_{\mu} \dif x^{\mu}=
      \frac{a}{g}\textbf{T}_3 + \frac{1}{g}(\im w\textbf{T}_1+\re w\textbf{T}_2)\dif\theta+\\
      \frac{1}{g}(\im w\textbf{T}_2-\re w\textbf{T}_1+\cot\negthinspace\theta\textbf{T}_3)\sin\negthinspace\theta\dif\phi
      +\frac{b}{2g}\lambda_8 \,.
      \label{witten}
\end{multline}
We already introduced an additional contribution in the simple 
$SU(3)$ case, where the $SU(2)$ generators are the trivially embedded
into the $SU(3)$ group, called isospin-$1/2$ embedding. 
There is a second, more involved, isospin-$1$ embedding of 
the $SU(2)$ group into the $SU(3)$ group which will not be used 
here~\cite{Corrigan:1976hd,Galtsov:1992au}.

Equation~(\ref{witten}) is invariant under $\mathrm{U}(1)$ gauge
transformations generated 
by $\mathrm{U} = \exp(i \Omega(t,r) \textbf{T}_3 )$,
under which 
\begin{align}
      a_a\mapsto a_a + \partial_a \Omega\,,\qquad
      w\mapsto e^{i \Omega} w\,,\qquad b \mapsto b\,.
      \label{g_trafo}
\end{align}

The four dimensional Yang-Mills action reads
\begin{align}
      L = -\int\frac{1}{2}\mbox{Tr}(F \Hodge F)\,,
\end{align}
where the gauge field strength is given by
$F=\dif A-ig[A,A]$ and $g$ is the gauge coupling constant.
This action is invariant under gauge transformations of $A$,
under which $A\mapsto UAU^{-1}+(1/g)U\dif U^{-1}$ and
$F\mapsto UFU^{-1}$.

Then the spherically reduced Yang-Mills action of the gauge 
field ansatz (\ref{witten}) reads~\cite{Volkov:1998cc}
\begin{align}
      L_{\rm YM}=4\pi\int\bigl(
      -\frac{1}{4g^2}f^2-\frac{1}{4g^2}\mathfrak{f}^2+
      \frac{1}{g^2}\frac{|Dw|^2}{\Phi^2}-\frac{1}{g^2}
      \frac{(|w|^2-1)^2}{2\Phi^4}\omega_g\bigr)\Phi^2\,,       
      \label{ymaction}
\end{align} 
where $f^2=f\Hodge f$, $|Dw|^2=|Dw\Hodge Dw|$ and the $2d$ abelian field 
strengths are given by
\begin{align}
      f&=\dif a \,,
      \label{2dfielda}\\
      \mathfrak{f}&=\dif b\,.
      \label{2dfieldb}
\end{align}
The gauge covariant derivative 
is $D=\dif-ia$, when acting on scalars. In case
of arbitrary magnetic and electric charge the Yang-Mills
equations imply $w=0$. On the other hand, $w\neq0$ only allows
a magnetic monopole with unit charge but still an 
arbitrary electric one~\cite{Volkov:1998cc}.

The field strength~(\ref{2dfieldb}) does not emerge from
the commutator of the gauge covariant derivative $D=\dif-ia$. 
Thus there is no coupling between $w$ and $b$. 
This can be understood from the fact 
that $\lambda_8$ in~(\ref{witten}) commutes with the generators
$\lambda_1,\ldots,\lambda_3$ of the $SU(2)$ Lie sub-algebra.

Nonetheless the gauge covariant derivative defined below, equation~(\ref{gcov}), 
when acting on spinors, has an contribution due to $b$.  

\section{The spherically symmetric Standard Model}\label{se:sm}

The aim of the last section was to review the three formalisms needed
for spherical reduction. This section uses these
to spherically reduce the remaining parts of the SM of
particle physics. Furthermore, with all the machinery already at
hand, we spherically reduce torsion generated by fermions, which yields
a four fermion interaction term.

Our procedure of spherical reduction can easily be extended to also
include the new terms of the recently proposed New Minimal
Standard Model~\cite{Davoudiasl:2004be}. For example, dark matter
necessitates the introduction of a new real scalar field,
which was already spherically reduced in example~\ref{ex_scalar},
equation~(\ref{scalarred}).

\subsection{Reduction of the $SU(2)$ Yang-Mills Dirac system}
The spherical reduction of the interaction term was often performed 
by an ansatz for the Dirac 
spinors~\cite{Klinkhamer:2001cp,Ratra:1988dp}. 
With our methods we show that an additional term appears that
may have been overlooked in previous calculations.

The interaction term is described by 
\begin{align}
        L = \int \Bigl( \bar{\Psi}^{\Lambda}_{\alpha} (\gamma^{\mu})^{\alpha}{}_{\beta} \, e_{\mu}^{a} (gA_{a})_{\Lambda \Delta} \Psi^{\Delta \beta} \Bigr) \omega_g\,,
        \label{yminter}
\end{align}
where $\Lambda$ and $\Delta$ denote the group
indices of the Yang-Mills field $A_a$, whereas the
indices $\alpha$ and $\beta$ describe the group
indices of the Dirac four-spinors $\Psi$.

Writing action~(\ref{yminter}) in the spinor formalism of section~\ref{subsec_red_fer}
yields
\begin{align}
        L  = \int \Bigl( 
        \sqrt{2}\bar{\psi}^{\Lambda A'} g A_{AA' \Lambda \Delta} \psi^{\Delta A}
        \Bigr) \omega_G\,,
        \label{yminter_sp}
\end{align}
since only left-handed fermions are interacting in the SM and where 
the minimal substitution
\begin{align}
      \nabla_{AA'} \mapsto \nabla_{AA'} - i g A_{AA' \Lambda \Delta}\,,
      \label{minimal}  
\end{align} 
was used. The Yang-Mills field in terms of spinor
components become
\begin{align}
      \iota^A \iota^{A'}A_{AA'}&=\frac{1}{g}(n^a a_a)\textbf{T}_3\,,
      \label{a00}\\ 
      o^A o^{A'}A_{AA'}&=\frac{1}{g}(l^a a_a)\textbf{T}_3\,,
      \label{a11}\\
      o^A \iota^{A'}A_{AA'}&=\frac{-i}{\sqrt{2}g \Phi} 
      \left(\bar{w} \textbf{T}_{-} -\cot\theta\, \textbf{T}_3 \right)\,, 
      \label{a10}\\
      \iota^A o^{A'}A_{AA'}&=\frac{i}{\sqrt{2}g \Phi}
      \left(w \textbf{T}_{+} -\cot\theta\, \textbf{T}_3 \right)\,,
      \label{a01}
\end{align}
where
\begin{align}
        \textbf{T}_{\pm} = \textbf{T}_1 \pm i \textbf{T}_2\,.
        \label{t+-}
\end{align}
In analogy to subsection~\ref{subsec_red_fer} the left-handed spinor 
$\psi^{\Delta A}$ is written as
\begin{align}
        \psi^{\Delta A} = A^{\Delta} o^A + P^{\Delta} \iota^A\,,
        \label{psi_rep}
\end{align}
where the two-component objects $P^{\Delta}$ and $A^{\Delta}$ are written as
\begin{align}
      P^{\Delta} = \begin{pmatrix} P^1 \\ P^2 \end{pmatrix}\,,\qquad
      A^{\Delta} = \begin{pmatrix} A^1 \\ A^2 \end{pmatrix}\,.
      \label{twocomp}
\end{align}
Putting in the components of the Yang-Mills fields 
in~(\ref{yminter_sp}) leads to
\begin{multline}
        L = \sqrt{2} \int \Bigl( 
        \bar{A} l^a a_a \textbf{T}_3 A + 
        \bar{P} n^a a_a \textbf{T}_3 P +
        \bar{P}\frac{-i}{\sqrt{2} \Phi} 
        \left(\bar{w}\textbf{T}_{-}-\cot\theta\, \textbf{T}_3 \right) A \\+
        \bar{A}\frac{i}{\sqrt{2} \Phi}
        \left(w\textbf{T}_{+}-\cot\theta\, \textbf{T}_3 \right) P
        \Bigr) \omega_G\,.
        \label{yminter3}
\end{multline}
Next we expand the functions $A^{\Delta}$ 
and $P^{\Delta}$ in terms of spin weighted 
spherical harmonics and integrate out the sphere.
To simplify the following calculations the spin weighted spherical 
harmonics are restricted to the cases $j=1/2$ and $m=\pm 1/2$
\begin{align}
        A^{\Delta} = \sum_{m= \pm \frac{1}{2}} 
        A^{\Delta}_{\frac{1}{2}\, m}\ {}_{-\frac{1}{2}} Y_{\frac{1}{2}\, m}\,,
        \qquad
        P^{\Delta} = \sum_{m= \pm \frac{1}{2}} 
        P^{\Delta}_{\frac{1}{2}\, m}\ {}_{\frac{1}{2}} Y_{\frac{1}{2}\, m}\,.
\end{align}
Spherical reduction of the first two terms yields
\begin{multline}
        L_{1} = \frac{\sqrt{2}}{2}\sum_{m= \pm \frac{1}{2}} \int \Bigl(
        \bar{A}^{1}_{\frac{1}{2}\, m} l^a a_a A^{1}_{\frac{1}{2}\, m}
        - \bar{A}^{2}_{\frac{1}{2}\, m} l^a a_a A^{2}_{\frac{1}{2}\, m} \\
        + \bar{P}^{1}_{\frac{1}{2}\, m} n^a a_a P^{1}_{\frac{1}{2}\, m}
        - \bar{P}^{2}_{\frac{1}{2}\, m} n^a a_a P^{2}_{\frac{1}{2}\, m}
        \Bigr) \Phi^2 \omega_{g}\,.
        \label{ym_red1}
\end{multline}
In the remaining two terms of~(\ref{yminter3}) 
the $\cot\negthinspace\theta$ terms vanish if the integration over 
the sphere is performed. $\textbf{T}_{+}$
and $\textbf{T}_{-}$ project out one component of
$A^{\Delta}$ and $P^{\Delta}$, respectively. 
Integrating out the sphere in the remaining
terms gives $\pm \pi/4$. Thus one finds
\begin{multline}
        L_{2} =i\frac{\pi}{4} \int \Bigl(       
        \bar{P}^{2}_{\frac{1}{2}\,\frac{1}{2}} \frac{\bar{w}}{\Phi} A^{1}_{\frac{1}{2}\,\frac{1}{2}}
        - \bar{P}^{2}_{\frac{1}{2}\,-\frac{1}{2}} \frac{\bar{w}}{\Phi} A^{1}_{\frac{1}{2}\,-\frac{1}{2}} \\
        - \bar{A}^{1}_{\frac{1}{2}\,\frac{1}{2} } \frac{w}{\Phi} P^{2}_{\frac{1}{2}\,\frac{1}{2}}
        + \bar{A}^{1}_{\frac{1}{2}\,-\frac{1}{2} } \frac{w}{\Phi} P^{2}_{\frac{1}{2}\,-\frac{1}{2}}
        \Bigr) \Phi^2 \omega_{g}\,,
        \label{ym_red2}
\end{multline}
where we needed the explicit form of the harmonics (c.f.~appendix~\ref{app_Y}), 
to evaluate the integral over the sphere.
$L_1 + L_2$ represent the spherically 
reduced $SU(2)$ Yang-Mills-Dirac interaction term.
 
\subsection*{Two-spinor representation of the $SU(2)$ interaction term}

Following the notation of the former sections the two spinors are
defined by
\begin{align}
      \Psi^{\I}_{jm}=\begin{pmatrix}A_{jm}\\0\end{pmatrix}\,,\qquad 
      \Psi^{\II}_{jm}=\begin{pmatrix}0\\P_{jm}\end{pmatrix}\,,
\end{align} 
where $B=Q=0$ was taken because only left-handed fermions
couple to $SU(2)$-Yang-Mills fields in the SM. Rewriting~(\ref{ym_red1}) in terms of these 
two-spinors leads to
\begin{align}
        L_{1}=\sum_{m=\pm\frac{1}{2}}\int\Bigl(
        \bar{\Psi}^{\I}_{\frac{1}{2}\,m}l^a a_a\textbf{T}_{3}\gamma^{-}_{\I}\Psi^{\I}_{\frac{1}{2}\,m}+
        \bar{\Psi}^{\II}_{\frac{1}{2}\,m}n^a a_a\textbf{T}_{3}\gamma^{+}_{\II}\Psi^{\II}_{\frac{1}{2}\,m}
        \Bigr)\Phi^2 \omega_{g}\,,
\end{align}
whereas for~(\ref{ym_red2}) we find 
\begin{align}
        L_{2}=\frac{\pi}{4}\sum_{m=\pm\frac{1}{2}}(-)^{\frac{1}{2}+m}\int \Bigl(        
        \bar{\Psi}^{\I}_{\frac{1}{2}\,m}\frac{w}{\Phi}
        \textbf{T}_{+}I^{-1}\bar{\Psi}^{\II}_{\frac{1}{2}\,m}+
        \bar{\Psi}^{\II}_{\frac{1}{2}\,m}\frac{\bar{w}}{\Phi}
        \textbf{T}_{-}I\bar{\Psi}^{\I}_{\frac{1}{2}\,m}
        \Bigr)\Phi^{2}\omega_{g}\,.
\end{align} 
Before fully writing out the reduced interaction term, we study the $SU(3)$ case.

\subsection{Reduction of the $SU(3)$ Yang-Mills Dirac system}

The spherical reduction of the interaction of fermions and 
$SU(3)$ Yang-Mills fields is very similar to the $SU(2)$ case.
The additional terms in~(\ref{witten}) are just 
\begin{align}
      \iota^{A}\iota^{A'}A_{A A'}=\frac{1}{2g}n^a b_a \lambda_8\,,\qquad
      o^{A}o^{A'}A_{A A'}=\frac{1}{2g}l^a b_a \lambda_8\,,
\end{align}
and all equations of the former subsections hold if $\textbf{T}_i$
denote the first three $SU(3)$ generators. This depends on the fact that we
only consider the simple isospin-$1/2$ embedding of the $SU(2)$ 
group into $SU(3)$. In the isospin-$1$ case~\cite{Galtsov:1992au} things change
considerably, since spacetime and group indices mix and hence
the spherical reduction is much more involved. However, the presented
procedure can be applied straightforwardly. 

$P^{\Lambda}$ is now a three-component object 
and the only additional term in the action reads
\begin{align}
      L=\frac{\sqrt{2}}{2}\int\Bigl(
      \bar{A}l^a b_a \lambda_8 A+\bar{P}n^a b_a \lambda_8 P 
      \Bigr)\omega_{G}\,,
\end{align}
where the sphere can be integrated out easily to give
\begin{align}
      L=\frac{\sqrt{2}}{2}\sum_{jm}\int\Bigl(
      \bar{A}_{jm}l^a b_a \lambda_8 A_{jm}+\bar{P}_{jm}n^a b_a \lambda_8 P_{jm} 
      \Bigr)\Phi^2\omega_{g}\,.
      \label{dym1}
\end{align}
Combining the left-handed part of~(\ref{actionred2}) with 
the reduced terms finally leads to the reduced Dirac-Yang-Mills action
\begin{multline}
        L_{\rm DYM} = \sum_{m=\pm\frac{1}{2}}\int\Bigl( 
        \bar{\Psi}^{\I}_{\frac{1}{2}\,m}i E_{+}^{a}\gamma^{+}_{\I}\D_a\Psi^{\I}_{\frac{1}{2}\,m}+
        \bar{\Psi}^{\II}_{\frac{1}{2}\,m}i E_{-}^{a}\gamma^{-}_{\II}\D_a\Psi^{\II}_{\frac{1}{2}\,m}\\
        +\frac{1}{\Phi}\bar{\Psi}^{\II}_{\frac{1}{2}\,m}(
        \gamma^{\star}+(-)^{\frac{1}{2}+m}\frac{\pi}{4}\bar{w}\textbf{T}_{-})I
        \Psi^{\I}_{\frac{1}{2}\,m}\\
        +\frac{1}{\Phi}\bar{\Psi}^{\I}_{\frac{1}{2}\,m}(
        \gamma^{\star}+(-)^{\frac{1}{2}+m}\frac{\pi}{4}w\textbf{T}_{+})I^{-1}
        \Psi^{\II}_{\frac{1}{2}\,m}
        \Bigr)\Phi^2 \omega_{g}\,,
        \label{actionDYM}
\end{multline}
with the gauge covariant derivative 
\begin{align}
        \D_a = \nabla_a - i a_a\textbf{T}_{3}-i \frac{b_a}{2} \lambda_8\,,
        \label{gcov}
\end{align}
when acting on fermions. In case of $SU(2)$ Yang-Mills theory, 
i.e., $b=0,\ \textbf{T}_i=\sigma_i/2$, the above action~(\ref{actionDYM}) 
is in agreement with references~\cite{Klinkhamer:2001cp,Ratra:1988dp} 
if only one value of the 'magnetic' quantum number $m$ is considered. 

Exact solutions of the $SU(2)$ and $SU(4)$ Einstein-Yang-Mills-Dirac systems
by reduction methods were found in~\cite{Rudolph:1997gz}. In addition to
spherical symmetry these authors also assumed homogeneity, hence
considered cosmological solutions. 

\subsection{The Higgs model}
In the action of the Higgs model one considers a complex scalar field with mass and 
self-interaction term
\begin{align}
      L=\int\bigl(
      G^{\mu\nu}(\D_{\mu}H)^{\dagger}\D_{\nu}H
      -\frac{\lambda}{4}(H^{\dagger}H-v^2)^2
      \bigr)\omega_{G}\,,
      \label{higgs_4d}
\end{align}
where the gauge covariant derivative 
reads $\D_{\mu}H=\nabla_{\mu}H-igA_{\mu}H$. In
spherical
symmetry~\cite{Greene:1993fw,Mondaini:1983jv,Ratra:1988dp,Volkov:1998cc} 
the Higgs field is given by\footnote{We stick with the notation of~\cite{Volkov:1998cc}.}
\begin{align}
      H=\frac{v}{g}\varphi\exp(i\xi\textbf{T}_{r})\ket{a}\,,
      \label{higgs_field}
\end{align}
where $\varphi=\varphi(x^{\alpha})$ and $\xi=\xi(x^{\alpha})$ are real
functions and $\ket{a}$ is a constant unit spinor, $\braket{a}{a}=1$.
The radial Pauli matrix $\textbf{T}_{r}$ is defined by
\begin{align}
      \textbf{T}_{r}=\sin\negthinspace\theta \cos\negthinspace\phi \textbf{T}_{1}
                    +\sin\negthinspace\theta \sin\negthinspace\phi \textbf{T}_{2}
                    +\cos\negthinspace\theta \textbf{T}_{3}\,.
      \label{eq:r_pauli}
\end{align}
Note that the Higgs field ansatz~(\ref{higgs_field}) differs from 
the standard parametrisation in particle physics. There the Higgs field
is usually parametrised by its shift around the vacuum expectation 
value $H\mapsto H_0 + H'$, where $H_0$ denotes the vacuum expectation value
and $H'$ is the shifted field.
The classical potential in~(\ref{higgs_4d}) vanishes for $H_0^{\dagger}H_0=v^2$.
Therefore one sees that the function $\varphi$ represents the
deviation around the minimum of the potential, but in contrast to the above, 
by multiplication rather than addition.

The exponential $\exp(i\xi\textbf{T}_{r})$ written explicitly yields
\begin{align}
      \exp(i\xi\textbf{T}_{r})=\cos\negthinspace\frac{\xi}{2}\unity
      +i\sin\negthinspace\frac{\xi}{2}
      \begin{pmatrix}
      \cos\negthinspace\theta & e^{-i\phi}\sin\negthinspace\theta \\
      e^{i\phi}\sin\negthinspace\theta & -\cos\negthinspace\theta
      \end{pmatrix}\,.
      \label{hedgehog}
\end{align}
The second term contains the spherical harmonics 
with $s=0$ and $l=1$ (see appendix~\ref{app_Y}). Before
performing the spherical reduction it should be noted that one could set $\xi=0$
and fix the isospin direction. The Higgs field~(\ref{higgs_field})
is still spherically symmetric is some sense but not spherically 
symmetric up to gauge transformations, so not according to our
second notion. When we analyse the effective theory
in two dimensions in section~\ref{se:4} we will choose the gauge 
$\xi=0$ to simplify the further calculation.

Spherical reduction of the Higgs action~(\ref{higgs_4d}) using
the ansatz~(\ref{higgs_field}) leads to
\begin{align}
      L_{\rm H}=\int\Bigl(\frac{v^2}{g^2} |Dh|^2 -\frac{v^2}{2g^2\Phi^2}\varphi^2 |w-e^{i\xi}|^2
      -\frac{\lambda v^4}{4g^4}(\varphi^2 -g^2)^2
      \Bigr)\Phi^2\omega_{g}\,,
      \label{red_higgs}
\end{align}
where $h=\varphi\exp(i\xi/2)$ and the gauge covariant derivative
reads $\D_{\alpha}=\nabla_{\alpha}-ia_{\alpha}/2$. (For $w$ recall
the remark after equation~(\ref{eym11})). The Higgs mass
and the vector boson mass are given by $M_H^2=2\lambda v^2$ and 
$M_W^2=g^2 v^2$, respectively.

The spherical reduction procedure yields an additional dilaton
dependent term in the Higgs potential. Hence the effective potential
in~(\ref{red_higgs}) reads
\begin{align}
      V = \frac{v^2}{g^2}\bigl(
      \frac{\varphi^2}{2 \Phi^2}|w-e^{i\xi}|^2 +\frac{\lambda v^2}{4g^2}(\varphi^2 -g^2)^2
      \bigr)\,,
\end{align}
which has a global minimum if 
\begin{align}
      \Phi \leq \sqrt{2}\frac{|w-e^{i\xi}|}{M_H}\,,
      \label{higgs_rad1}
\end{align}
and its usual symmetry breaking form otherwise. 
Following e.g.~\cite{Greene:1993fw} finite energy solutions
require $|w|^2<1$ and hence $|w-e^{i\xi}|<2$.
Therefore from~(\ref{higgs_rad1}) we conclude
\begin{align}
      \Phi \leq \frac{2\sqrt{2}}{M_H}\,.
      \label{higgs_rad2}
\end{align}
For $M_H\approx 100 {\rm GeV} - 1 {\rm TeV}$ this yields a radius of ca.~$10^{16}-10^{17}$ Planck lengths.

Restoration of symmetry at some small
radius, e.g. near a black hole horizon~\cite{Hawking:1980ng,Page:1982fm}, 
can be understood from the following consideration. 
An observer would not see the Higgs field in some 
vacuum state but rather in a thermal bath of Hawking 
quanta close to a black hole. Hence, if that temperature is high 
enough, the potential smears out.

\subsection{Yukawa couplings}
In the SM of particle physics fermion masses are introduced by
Yukawa couplings and the Higgs mechanism. 
Therefore the explicit mass terms in the fermion sector,
section~\ref{subsec_red_fer}, can be ignored henceforth.
The Yukawa interaction term reads
\begin{align}
      L=\int\chi_{A'}\bar{\psi}^{A'\Delta}H_{\Delta}\omega_{G}\,,
      \label{yukawa}
\end{align}
where the internal group index in the Higgs field indicates
the presence of the unit spinor $\ket{a}$ in~(\ref{higgs_field}).
Since the action~(\ref{yukawa}) is not hermitian one must add
the hermitian conjugate, which we do at the end.

Following the procedure of the previous sections we first
write the spinors $\chi_{A'}$ and $\bar{\psi}^{A'\Delta}$ in terms 
of basis spinors and get
\begin{align}
      L=\int\Bigl(\frac{v}{g}\varphi(Q\bar{P}^{\Delta}-B\bar{A}^{\Delta})
      \exp(i\xi\textbf{T}_{r})\ket{a}
      \Bigr)\Phi^2\omega_{g}\dif^2\Omega\,.
\end{align}
Next we expand the coefficients
$A,B,P,Q$ in terms of spin weighted spherical harmonics and use
their explicit form~(\ref{hedgehog}). Furthermore, to  simplify the 
following, we choose the unit vector to be $\bra{a}=(0,\,1)$. 
The first term of~(\ref{hedgehog}) is easily reduced because 
one can use the orthogonality condition~(\ref{ortho}) since $A,B$ 
and $P,Q$ have the same spin weights respectively. This yields
\begin{align}
      L_1=\sum_{m=\pm\frac{1}{2}}\int\Bigl(\frac{v}{g}\varphi\cos\negthinspace\frac{\xi}{2}
      (Q_{\frac{1}{2}\,m}\bar{P}_{\frac{1}{2}\,m}^{2}-
      B_{\frac{1}{2}\,m}\bar{A}_{\frac{1}{2}\,m}^{2})
      \Bigr)\Phi^2\omega_{g}\,.
\end{align}
The second and more involved term after spherical reduction reads
\begin{multline}
      L_2=\int \frac{v}{g}\varphi i\sin\negthinspace\frac{\xi}{2}\Bigl(
      \frac{2}{3}(\bar{A}^1_{\frac{1}{2}\,-\frac{1}{2}}B_{\frac{1}{2}\,\frac{1}{2}}-
      \bar{P}^1_{\frac{1}{2}\,-\frac{1}{2}}Q_{\frac{1}{2}\,\frac{1}{2}})\\+
      \frac{1}{3}(\bar{A}^2_{\frac{1}{2}\,\frac{1}{2}}B_{\frac{1}{2}\,\frac{1}{2}}-
      \bar{A}^2_{\frac{1}{2}\,-\frac{1}{2}}B_{\frac{1}{2}\,-\frac{1}{2}}-
      \bar{P}^2_{\frac{1}{2}\,\frac{1}{2}}Q_{\frac{1}{2}\,\frac{1}{2}}+
      \bar{P}^2_{\frac{1}{2}\,-\frac{1}{2}}Q_{\frac{1}{2}\,-\frac{1}{2}})
      \Bigr)\Phi^2\omega_{g}\,.
\end{multline}
The last two terms plus their hermitian 
conjugates will be written in terms of two-spinors.
The latter are defined by~(\ref{psiplus}) and~(\ref{psiminus}),
moreover~(\ref{twocomp}) is taken into account. Since we wish to
write the effective $2d$ action without the unit spinor $\ket{a}$,
the expansion coefficients $B$ and $Q$ are embedded in the complex two-spinor
space by
\begin{align}
      B_{jm} = \begin{pmatrix} 0\\ B_{jm} \end{pmatrix}\,,\qquad
      Q_{jm} = \begin{pmatrix} 0\\ Q_{jm} \end{pmatrix}\,.
\end{align}
Then the spherically reduced Yukawa term becomes
\begin{multline}
      L_{\rm Y}=\sum_{m=\pm\frac{1}{2}}\int\frac{v}{g}\varphi\Bigl(
      -\cos\negthinspace\frac{\xi}{2}
      (\bar{\Psi}^{\I}_{\frac{1}{2}\,m}\unity\Psi^{\I}_{\frac{1}{2}\,m}+
      \bar{\Psi}^{\II}_{\frac{1}{2}\,m}\unity\Psi^{\II}_{\frac{1}{2}\,m})\\+
      i\frac{1}{3}\sin\negthinspace\frac{\xi}{2}(-)^{\frac{1}{2}+m}
      (\bar{\Psi}^{\I}_{\frac{1}{2}\,m}\gamma^{\star}\Psi^{\I}_{\frac{1}{2}\,m}-
      \bar{\Psi}^{\II}_{\frac{1}{2}\,m}\gamma^{\star}\Psi^{\II}_{\frac{1}{2}\,m})\\+
      i\frac{2}{3}\sin\negthinspace\frac{\xi}{2}
      (\bar{\Psi}^{\I}_{\frac{1}{2}\,-\frac{1}{2}}\textbf{T}_{+}
      P_{-}\Psi^{\I}_{\frac{1}{2}\,\frac{1}{2}}-
      \bar{\Psi}^{\I}_{\frac{1}{2}\,\frac{1}{2}}\textbf{T}_{-}
      P_{+}\Psi^{\I}_{\frac{1}{2}\,-\frac{1}{2}}\\+
      \bar{\Psi}^{\II}_{\frac{1}{2}\,-\frac{1}{2}}\textbf{T}_{+}
      P_{+}\Psi^{\II}_{\frac{1}{2}\,\frac{1}{2}}-
      \bar{\Psi}^{\II}_{\frac{1}{2}\,\frac{1}{2}}\textbf{T}_{-}
      P_{-}\Psi^{\II}_{\frac{1}{2}\,-\frac{1}{2}})
      \Bigr)\Phi^2\omega_{g}\,,
      \label{yukawa_xi}
\end{multline}
where we added the hermitian conjugate. $P_{\pm}$ are the usual
chiral projection operators defined by
\begin{align}
      P_{\pm}=\frac{1}{2}(\unity\pm\gamma^{\star})\,,
      \label{proj}
\end{align}
which are needed since left- and right-handed fermions
are coupled together. 

For later use we set $\xi=0$ in the spherically reduced Yukawa
action~(\ref{yukawa_xi}) which fixes the isospin direction. Then the
last three lines vanish and one is left with the simple term
\begin{align}
      L_{\rm Y}=\sum_{m=\pm\frac{1}{2}}\int
      \bigl(-\frac{v}{g}\varphi\bigr)
      \Bigl(\bar{\Psi}^{\I}_{\frac{1}{2}\,m}\unity\Psi^{\I}_{\frac{1}{2}\,m}+
      \bar{\Psi}^{\II}_{\frac{1}{2}\,m}\unity\Psi^{\II}_{\frac{1}{2}\,m}
      \Bigr)\Phi^2\omega_{g}\,.
      \label{yukawa_null}
\end{align}
Therefore the induced mass of the Yukawa coupling reads
\begin{align} 
      m_{\rm Y} = \frac{v}{g}\varphi\,,
      \label{yukawa_mass}
\end{align}
by comparison with the spherically reduced Dirac action~(\ref{actionred2}).
A small consistency check is to note that the negative sign of~(\ref{yukawa_null}) 
is consistent with~(\ref{actionred2}).

\subsection{Einstein-Cartan theory}
\label{ectheory}

As in section~\ref{subsec_cartan} torsion is most naturally included by assuming the existence 
of a derivative operator $\tilde{\nabla}_a$ that is not
torsion-free. That derivative operator can be split into a torsion-free 
part $\nabla_a$ and a torsion dependent part by
\begin{align}
      \tilde{\nabla}_a U^c = \nabla_a U^c +K_{ab}{}^c U^b\,,
      \label{cov}
\end{align}  
where $K_{ab}{}^c$ is called the contortion tensor and where we 
follow the notation of Penrose~\cite{penrosespinors}.
$K_{ab}{}^c$ is the holonomic version of~(\ref{con_form})
with an additional negative sign because of the different index
positions used in the different formalisms.

Metricity of both covariant derivative operators immediately
implies antisymmetry of $K_{ab}{}^c$ in the last index pair.
Contortion and torsion are related by
\begin{align}
      \tilde{T}_{ab}{}^c = K_{ba}{}^c - K_{ab}{}^c\,. 
\end{align}
When $\tilde{\nabla}_a$ is acting on spinors we write
\begin{align}
      \tilde{\nabla}_{AA'}\psi^C&=\nabla_{AA'}\psi^C+ 
      \Theta_{AA'B}{}^C \psi^B,
      \label{covspin1}\\
      \tilde{\nabla}_{AA'}\chi^{C'}&=\nabla_{AA'}\chi^{C'}+
      \bar{\Theta}_{AA'B'}{}^{C'} \chi^{B'}\,,
      \label{covspin2}
\end{align}
from which the contortion tensor can be reconstructed when 
the action on a vector $U^c=U^{CC'}$
\begin{align} 
      K_{ab}{}^c = \Theta_{AA'B}{}^{C}\e_{B'}{}^{C'}
      + \bar{\Theta}_{AA'B'}{}^{C'} \e_B{}^C\,,
      \label{ktheta}
\end{align}
is considered. Torsion can now be incorporated in the former equations
by replacing $\nabla_a$ by $\tilde{\nabla}_a$. Then one 
uses~(\ref{cov}) and~(\ref{covspin1}), (\ref{covspin2})
and finds additional contributions containing the
contortion spinor. The latter can
be decomposed further into irreducible parts by 
\begin{align}
      \Theta_{A'ABC} = \Theta_{A'(ABC)} +\frac{1}{3}\e_{AB}\Theta_{A'C}
      +\frac{1}{3}\e_{AC}\Theta_{A'B}\,,
      \label{theta_dec}
\end{align}
where the trace terms are  
\begin{align} 
      \Theta_{A'B}=\Theta_{DA'B}{}^{D},\qquad
      \bar{\Theta}_{AB'}=\bar{\Theta}_{D'AB'}{}^{D'}\,.
      \label{theta_trace}
\end{align}
Using the above, a third way of writing the Einstein-Hilbert-Cartan action is
\begin{align}
      L_{\rm EHC}&=\int\tilde{R}\, \omega_G = 
      \int \Bigl(R+K_{ae}{}^b K_{b}{}^{ae}-
      K_{be}{}^b K_{a}{}^{ae}\Bigr) \omega_{G}\nonumber\\
      &=\int \Bigl(
      R + \frac{4}{3}\Theta_{A'B}\Theta^{A'B} + 
      \frac{4}{3}\bar{\Theta}_{AB'}\bar{\Theta^{A'B}} \nonumber\\
      &\phantom{=\int \Bigl(}
      -\Theta_{A'(ABC)}\Theta^{A'(CAB)} - 
      \bar{\Theta}_{A(A'B'C')} \bar{\Theta}^{A(C'A'B')}
      \Bigr) \omega_{G}\,,
      \label{hecaction} 
\end{align}
where the surface term is omitted, see e.g.~\cite{Zecca:2002de}. 
The introduction of $\tilde{\nabla}_a$ in the Dirac
action functional~(\ref{action}) leads to
an additional term 
\begin{align}
      L_{\rm DT}=\frac{i}{\sqrt{2}} \int \Bigl(
      \Theta_{A'B}(\bar{\psi}^{A'}\psi^B-\chi^{A'}\bar{\chi}^B) -
      \bar{\Theta}_{AB'}(\psi^A\bar{\psi}^{B'}-\bar{\chi}^A\chi^{B'} )
      \Bigr) \omega_G\,.
      \label{dirac_dt}
\end{align}
From~(\ref{hecaction}) and~(\ref{dirac_dt}) one can derive the equations of motion
for the contortion contribution. Variation with respect to
$\delta \Theta_{A'(ABC)}$ yields the trivial equation of 
motion $\Theta_{A'(ABC)}=0$. Variation with respect to $\delta \Theta_{A'B}$ yields
\begin{align}
      \frac{\delta L_{EHT}}{\delta \Theta_{A'B}} = \frac{8}{3} \Theta^{A'B}\,,\qquad
      \frac{\delta L_{DT}}{\delta \Theta_{A'B}} = \frac{i}{\sqrt{2}} 
      (\bar{\psi}^{A'}\psi^B-\chi^{A'}\bar{\chi}^B)\,,
\end{align}
which implies an algebraic equation of motion
\begin{align}
      \Theta^{A'B} = i\frac{3}{8\sqrt{2}}
      (\bar{\psi}^{A'}\psi^B-\chi^{A'}\bar{\chi}^B)\,,
      \label{contorsion_eom}
\end{align}
for the trace of the contortion spinor. We already argued
in subsection~\ref{subsub_hilbert} that this is expected
on general grounds. The contortion spinor is given 
by the fermion current and is purely imaginary. 

In the literature~\cite{Alimohammadi:1998vx,Griffiths:1981ym,Hehl:1973} the 
statement is often found that
Dirac fermions only couple to the axial torsion vector or
that the contortion tensor is totally skew-symmetric. 
This can easily be understood in
the spinor formalism since one easily checks that
\begin{align}
      A^a = \frac{2}{3}\im\, \Theta^{AA'}\,,\qquad
      k^a = -\Phi 2\re\, \Theta^{AA'}\,,
\end{align}
where $A^a$ and $k^a$ are the holonomic, not yet spherically
reduced versions of~(\ref{vector}). The vector $A^a$ given by
the equation of motion for contortion clearly has components along
the $m^a\,,\bar{m}^a$ directions.
The minus sign in $k^a$ is due to the different conventions,
already mentioned in the beginning of this subsection.
Since fermions only couple to the axial contortion vector
variation with respect to $k^a$ and $U_{lmn}$ must vanish. 
The vanishing of $U_{lmn}$ implies that 
\begin{align}
      s_a = \frac{1}{2\Phi} h_a\,,
\end{align}
as can be seen from~(\ref{u2}). 
Since the equation of motion~(\ref{contorsion_eom}) is purely 
algebraic one can eliminate the contortion terms 
from $L_{{\rm EHT}}$ and $L_{{\rm DT}}$. Since
\begin{align}
      \frac{4}{3}\Theta^{A'B}\Theta_{A'B} &= 
      \frac{3}{16}(\bar{\psi}^{A'} \psi^B \chi_{A'} \bar{\chi}_B)\,, \\
      \frac{i}{\sqrt{2}}\Theta_{A'B}(\bar{\psi}^{A'} \psi^B -\chi^{A'} \bar{\chi}^B) &=
      \frac{3}{8}(\bar{\psi}^{A'} \psi^B \chi_{A'} \bar{\chi}_B )\,,
\end{align}
one finds that the elimination yields $\tau=2(3/16+3/8)=9/8$
\begin{align}
      L_{\rm T} = L_{\rm EHC} + L_{\rm DT} =\tau \int
      (\bar{\psi}^{A'} \chi_{A'} \psi^B \bar{\chi}_B) 
      \omega_G\,,
      \label{action_t}
\end{align}
an effectively four-fermion interaction term. It has the structure of a
dilaton deformed Thirring model~\cite{Thirring:1958in}. If the fermion action only 
consists of chiral fermions, then either $\psi^A$ of $\chi^{A'}$ is
zero, hence the action~(\ref{action_t}) would vanish. Therefore 
torsion generated by fermions is nontrivial if and only if both
four dimensional chiralities are present. However, only one
of the invariant two-spinors~(\ref{psiplus}) or~(\ref{psiminus}) 
is needed to generate torsion. 

If the Dirac two-spinors are expanded in terms of basis
spinors the action~(\ref{action_t}) becomes
\begin{align}
      L_{\rm T} = \tau\int(\bar{P}Q-\bar{A}P)(P\bar{Q}-A\bar{B})\omega_G\,.
      \label{action_t2}
\end{align}
As already pointed out the standard model with torsion is
characterised by one additional term only, namely the four-fermion
interaction term~(\ref{action_t}). This action can be spherically
reduced by the above methods. We expand the functions $A,B,P,Q$ 
in terms of spin weighted spherical harmonics ${}_s Y_{jm}$ with 
the additional restriction $j=1/2$ and $m=\pm 1/2$. 

Putting the expansion in the action~(\ref{action_t2}) 
gives $2^6 = 64$ terms. Next the spherical dependence can be integrated
out and one has to evaluate inner products of four spin weighted spherical
harmonics. $40$ of these inner product vanish and one is left with 
$24$ non-vanishing terms, which equals the number of independent
components of the torsion or contortion tensor. Note that these 24
non-vanishing terms are not independent since for fermions 
the contortion tensor has only four independent components. 
The inner products are given in 
appendix~\ref{innerproducts}. As before these terms can be written
in terms of two spinors~(\ref{psiplus}), (\ref{psiminus}), which are
also given in appendix~\ref{innerproducts}. 

For the moment we put $\Psi^{\II}_{jm}=0$ and moreover assume that only one
'magnetic' quantum number $m$ is present, say $m=1/2$, in~(\ref{d1}) 
and~(\ref{d2}). Then the simplest non-trivial torsion term
becomes
\begin{align}   
      L_{\rm T}=\frac{\tau}{3\pi}\int\Bigl(
      \bar{\Psi}^{\I}_{\frac{1}{2}\,\frac{1}{2}}P_{+}
      \Psi^{\I}_{\frac{1}{2}\,\frac{1}{2}})
      (\bar{\Psi}^{\I}_{\frac{1}{2}\,\frac{1}{2}}P_{-}
      \Psi^{\I}_{\frac{1}{2}\,\frac{1}{2}}
      \Bigr)\Phi^2 \omega_{g}\,,
\end{align}
which we simply state to show the general structure of those terms.

Similar to the Yukawa coupling, the projectors~(\ref{proj}) are needed 
since left- and right-handed fermions couple together. The factor $1/3\pi$
enters because of the integration of four spin-weighted spherical
harmonics.

For sake of completeness we mention some additional aspects of 
Einstein-Cartan theory. The GHP spin-coefficient 
formalism can be extended to include torsion~\cite{Griffiths:1980}.
The idea is based on equation~(\ref{cov}), one splits every
spin-coefficient into two parts, $\rho^0$ which is torsion free and
$\rho_1$ that depends on the contortion, where we adopted
the notation of~\cite{Griffiths:1980}. Therefore the complete
spin-coefficient is just the sum of those two parts.
The extended formalism was then used in~\cite{Griffiths:1981ym} to
analyse neutrino fields in Einstein-Cartan theory. There
the torsion spin-coefficients are
\begin{align}
      \rho_1 = ik \bar{\phi}\phi\,,\qquad \gamma_1 = \frac{ik}{2}\bar{\phi}\phi\,,
      \label{eq:rho_tor1}
\end{align}
where $\phi^A=\phi o^A$ is the neutrino field and $k$ is the coupling 
constant. Furthermore~\cite{Griffiths:1981ym} contains some interesting theorems, 
that we can use directly. One of them states (see \S(7) of~\cite{Griffiths:1981ym})
that ghost neutrinos, which have vanishing canonical energy-momentum
tensor, cannot be constructed in spherically symmetric spacetimes.

The perfect fluid considered previously in subsection~\ref{se:sspf}
can be generalised by a Weyssenhoff fluid~\cite{Weyssenhoff:1947}
which permits a non-vanishing spin density. It is characterised
by a classical description of spin, where the source term of
torsion is written $s^{\kappa}{}_{\mu\nu} = u^{\kappa}S_{\mu\nu}$,
with $u^{\kappa}$ the fluid's four velocity and $S_{\mu\nu}$
the intrinsic angular momentum tensor. In~\cite{Griffiths:1982}
a Weyssenhoff fluid determining torsion by one function $S$ was considered
within the framework of the extended spin-coefficient formalism by
the same authors. It was found that the torsion spin-coefficient are
\begin{align}
      \rho_1=-\rho'_1=2\gamma_1=-2\gamma'_1=iS\,.
      \label{eq:rho_tor2}
\end{align}
By considering a static and spherically symmetric Weyssenhoff
fluid in a cosmological context, one of the present authors 
could suggest a mechanism to solve the sign problem of the 
cosmological constant~\cite{Boehmer:2003iv}. Note that 
in equation~(\ref{eq:rho_tor1})
the torsion coefficient $\rho_1$ appears, although the torsion
free part $\rho_0$ drops out of the Dirac action~(\ref{action2}).

Einstein-Cartan theory is derived from the usual Einstein-Hilbert
action without restricting to torsionless spacetimes. This action
is linear in curvature. However, the term 
$\eps^{klmn}\tilde{R}_{klmn}$ is also linear in curvature 
and was considered in~\cite{Hojman:1980kv}. In case of vanishing
torsion this term identically vanishes which was already mentioned
in~\cite{Weinberg:1972}. If torsion is present then the term
gives additional contributions to the field equations. 
In~\cite{Hojman:1980kv} it was used to analyse parity violating
contributions in the action. Fortunately it turns out that this
term also vanishes identically if fermions are the source
of torsion as in the SM under consideration.
The first non-trivial contributions enters the field equations
if massive spin one particles are allowed to generate torsion,
which is beyond the scope of the present work.

\section{Effective theory in d=2}
\label{se:4}

The advantage of an effective theory in lower dimensions is twofold: at
the level of equations of motion the theory is equivalent to the
higherdimensional one, but the classical analysis is much simpler. Thus,
exact solutions can be constructed with particular ease. However, there
is more to the lowerdimensional theory than just a convenient scheme for
reproduction: it can be treated as a model on its own and semi-classical
and quantum aspects can be studied in detail. This can provide valuable
insight into the quantum regime of the higherdimensional theory,
although one has to be careful with interpreting the results because
spherical reduction and renormalisation need not to commute~\cite{Frolov:1999an}. 

We now present the SSSMG as an effective $2d$ theory and address its
quantisation.

\subsection{The SSSMG as a $2d$ model}

We now combine the spherically reduced actions of the former sections
into an effectively $2d$ action which represents the 
SSSMG in first order form
\begin{align}
      L=L_{\rm FOG}+L_{{\rm YM}}^{U(1)}+L_{{\rm YM}}^{SU(2)}+L_{{\rm YM}}^{SU(3)}
         +L_{\rm DYM}+L_{\rm H}+L_{\rm Y}+L_{\rm T}\,,
      \label{eq:5.1}
\end{align}
where the different parts of the action are given by ($\epsilon:=e^+\wedge e^-$)
\begin{align}
      L_{\rm FOG}&=\frac{2\pi}{\la^2}\int_L \Bigl(X_a (D\wedge e)^a 
      +X\dif\omega + \epsilon{\mathcal V} (X,X^aX_a) \Bigr)\,,\\
      L_{{\rm YM}}^{U(1)}&=\frac{4\pi}{g^2_1}\int_L
      \Bigl(z_1\dif a_1 +\epsilon \frac{z^2_1}{X}\Bigr)\,,\\
      L_{{\rm YM}}^{SU(2)}&=\frac{4\pi}{g^2_2}\int_L\Bigl(z_2\dif a_2 +\epsilon \frac{z^2_2}{X}+
      |\D w_2 \wedge \ast (\D w_2)|-\frac{(|w_2|^2-1)^2}{2X}\epsilon\Bigr)\,,\\       
      L_{{\rm YM}}^{SU(3)}&=\frac{4\pi}{g^2_3}\int_L\Bigl(z_3\dif a_3+y\dif b 
      +\epsilon \frac{z^2_3+y^2}{X}\nonumber \\
      &\mbox{\hspace{1cm}} +|\D w_3 \wedge \ast (\D w_3)|-\frac{(|w_3|^2-1)^2}{2X}\epsilon\Bigr)\,,\\
      L_{\rm DYM}&= \sum_{m=\pm\frac12} \int_L \Biggl( X\sum_{N=\I,\II} 
      \bigl(i\bar{\Psi}^N_{\frac12 m}\ga_N^a e_a\wedge \ast D\Psi^N_{\frac12 m}\bigr) 
      \nonumber \\ 
      &\mbox{\hspace{1cm}}+(j+\frac{1}{2})\sqrt{X} \epsilon\Bigl(\bar{\Psi}^{\II}_{\frac12m}\bigl(
      \gamma^{\star}+(-)^{\frac{1}{2}+m}\frac{\pi}{4}\bar{w}\textbf{T}_{-}\bigr)I\Psi^{\I}_{\frac12 m}
      \nonumber \\
      &\mbox{\hspace{2cm}}+\bar{\Psi}^{\I}_{\frac12m}\bigl(\gamma^{\star}+(-)^{\frac{1}{2}+m}
      \frac{\pi}{4}w\textbf{T}_{+}\bigr)I^{-1}
      \Psi^{\II}_{\frac12 m}\Bigr)\Biggr)\,,
\end{align}
\begin{align}
      L_{\rm H}&=\frac{v^2}{g^2_2}\int_L\Bigl(X|Dh\wedge \ast Dh|-\frac{1}{2}\epsilon\varphi^2 
      |w_2-e^{i\xi}|^2
      -X\epsilon\frac{\lambda v^2}{4g^2_2}(\varphi^2 -g^2)^2
      \Bigr)\,.
\end{align}
New auxiliary fields $y,z_i$ have been introduced in order to bring the 
Yang-Mills part (with gauge field 1-forms $b,a_i$) into first order form. 
For convenience of the reader we recall the field content:
$(X,X^a,z_1,z_2,z_3,y)$ are scalar fields which in the absence of matter
can be interpreted as target space coordinated of a Poisson manifold~\cite{Schaller:1994es};
$(\omega,e^a,a_1,a_2,a_3,b)$ are connection, zweibein, $U(1)$ connection,
$SU(2)$ connection, $SU(3)$ connection, respectively. $w_2$ and $w_3$ are
complex scalar fields coming from the reduction of the $SU(2)$ and 
$SU(3)$ connection, respectively. The complex scalar
$h=\varphi\exp(i\xi/2)$ is the Higgs field. 
$\Psi$ represents {\em all} SM fermions. 

The gauge covariant derivatives read
\begin{align}
      (\D \wedge e)^\pm &= \dif e^\pm \pm \om\wedge e^\pm\,,\\
      \D w_2 &=\dif w_2-ia_2 w_2\,,\\
      \D w_3 &=\dif w_3-ia_3 w_3\,,\\
      \mbox{{\small neutrinos:}\ } \D \Psi &=
      \dif\Psi-i\frac{a_2}{2}\sigma_{3}\Psi \,,
      \label{eq:der_neu} \\
      \mbox{{\small charged leptons:}\ } \D \Psi &=
      \dif\Psi-ia_1\Psi-i\frac{a_2}{2}\sigma_{3}\Psi \,,
      \label{eq:der_lep} \\
      \mbox{{\small quarks:}\ } \D \Psi &= 
      \dif\Psi-ia_1\Psi-i\frac{a_2}{2}\sigma_{3}\Psi
      -i\frac{a_3}{2}\lambda_{3}\Psi-i\frac{b}{2}\lambda_8\Psi\,,
      \label{eq:der_qua} \\
      \D h &=\dif h-i\frac{a_2}{2} h\,.
\end{align}
(\ref{eq:der_neu})--(\ref{eq:der_qua}) cover all covariant derivatives
with and without $SU(2)$ couplings.  

Different parts of the above actions and permutations
thereof were a rich source of analytical and numerical investigations
during the last 15 years. The likely starting point was~\cite{Bartnik:1988am}. 
In recent years the inclusion of the
cosmological constant with its non-flat asymptotic structure motivated
further studies of the above system, going back probably 
to~\cite{Torii:1995wv}. 

The remaining terms of the standard model are
\begin{multline}
      L_{\rm Y}=\sum_{m=\pm\frac{1}{2}}\int_L\frac{v}{g_2}\varphi\Bigl(
      -\cos\negthinspace\frac{\xi}{2}
      (\bar{\Psi}^{\I}_{\frac{1}{2}\,m}\unity\Psi^{\I}_{\frac{1}{2}\,m}+
      \bar{\Psi}^{\II}_{\frac{1}{2}\,m}\unity\Psi^{\II}_{\frac{1}{2}\,m})\\
      +i\frac{1}{3}\sin\negthinspace\frac{\xi}{2}(-)^{\frac{1}{2}+m}
      (\bar{\Psi}^{\I}_{\frac{1}{2}\,m}\gamma^{\star}\Psi^{\I}_{\frac{1}{2}\,m}-
      \bar{\Psi}^{\II}_{\frac{1}{2}\,m}\gamma^{\star}\Psi^{\II}_{\frac{1}{2}\,m})
      \phantom{XXXX}\\
      +i\frac{2}{3}\sin\negthinspace\frac{\xi}{2}
      (\bar{\Psi}^{\I}_{\frac{1}{2}\,-\frac{1}{2}}\textbf{T}_{+}
      P_{-}\Psi^{\I}_{\frac{1}{2}\,\frac{1}{2}}-
      \bar{\Psi}^{\I}_{\frac{1}{2}\,\frac{1}{2}}\textbf{T}_{-}
      P_{+}\Psi^{\I}_{\frac{1}{2}\,-\frac{1}{2}}\\
      +\bar{\Psi}^{\II}_{\frac{1}{2}\,-\frac{1}{2}}\textbf{T}_{+}
      P_{+}\Psi^{\II}_{\frac{1}{2}\,\frac{1}{2}}-
      \bar{\Psi}^{\II}_{\frac{1}{2}\,\frac{1}{2}}\textbf{T}_{-}
      P_{-}\Psi^{\II}_{\frac{1}{2}\,-\frac{1}{2}})
      \Bigr)X\epsilon\,.
\end{multline}
Finally the torsion terms are
\begin{multline}
      L_{\rm T} =   
      \frac{\tau}{3 \pi}\sum_{m=\pm\frac{1}{2}}\int_L\Bigl(
      (\bar{\Psi}^{\III}_{\frac{1}{2}\,\pm m}P_{+}
      \Psi^{\III}_{\frac{1}{2}\,\pm m})
      (\bar{\Psi}^{\III}_{\frac{1}{2}\,\pm m}P_{-}
      \Psi^{\III}_{\frac{1}{2}\,\pm m})\\-
      (\bar{\Psi}^{\I}_{\frac{1}{2}\,\mp m}P_{\pm}
      \Psi^{\I}_{\frac{1}{2}\,\mp m})
      (\bar{\Psi}^{\II}_{\frac{1}{2}\,\pm m}P_{\pm}
      \Psi^{\II}_{\frac{1}{2}\,\pm m})+
      \frac{1}{2}
      (\bar{\Psi}^{\III}_{\frac{1}{2}\,\mp m}P_{+}
      \Psi^{\III}_{\frac{1}{2}\,\mp m})
      (\bar{\Psi}^{\III}_{\frac{1}{2}\,\pm m}P_{-}
      \Psi^{\III}_{\frac{1}{2}\,\pm m})\\[1ex]+
      \frac{1}{2}
      (\bar{\Psi}^{\III}_{\frac{1}{2}\,\mp m}P_{+}
      \Psi^{\III}_{\frac{1}{2}\,\pm m})
      (\bar{\Psi}^{\III}_{\frac{1}{2}\,\pm m}P_{-}
      \Psi^{\III}_{\frac{1}{2}\,\mp m})- 
      \frac{1}{2}
      (\bar{\Psi}^{\I}_{\frac{1}{2}\,\pm m}P_{\pm}
      \Psi^{\I}_{\frac{1}{2}\,\pm m})
      (\bar{\Psi}^{\II}_{\frac{1}{2}\,\pm m}P_{\pm}
      \Psi^{\II}_{\frac{1}{2}\,\pm m})\\[1ex]+
      \frac{1}{2}
      (\bar{\Psi}^{\I}_{\frac{1}{2}\,\pm m}P_{\pm}
      \Psi^{\I}_{\frac{1}{2}\,\mp m})
      (\bar{\Psi}^{\II}_{\frac{1}{2}\,\pm m}P_{\pm}
      \Psi^{\II}_{\frac{1}{2}\,\mp m})
      \Bigr)X\epsilon\,.
\end{multline}

In case of Riemannian manifolds rather that Lorentzian ones, the
action of the torsion free standard model with gravity was
formally expressed in terms of Dirac-Yukawa operators in~\cite{Tolksdorf:1998ns}.

\newcommand{\extd}{\dif}
As an illustration why the reformulation as a 2D model is useful 
we consider now its classical solutions. Surprisingly, up to the 
very last step the construction of geometry works {\em exactly} 
as for the matterless case and the relevant details have been 
spelled out in section \ref{se:3.2}. So assuming that either $X^+$ 
or $X^-$ are non-vanishing in an open region the line element reads
\begin{align}
      \label{eq:lineelementursuper}
      \extd s^2=2\extd u\extd r + 2X^+X^-\extd u^2 - r^2\extd\Omega^2\,,
\end{align}
where $r\propto\sqrt{X}$ and the product $X^+X^-$ fulfils the 
conservation equation~(\ref{eq:pfe}). Of course, in general it 
is quite hopeless to integrate that equation which contains the 
information of {\em all} matter contributions described above; 
nevertheless, the simplicity of~(\ref{eq:lineelementursuper}) allows 
for some general statements, independent from material details: 
Apparent horizons are encountered for $X^+X^-=0$. If both $X^+=X^-=0$ 
at an isolated point the region around that point behaves like 
the region around the bifurcation 2-sphere of the Schwarzschild BH; 
in that case instead of (\ref{eq:lineelementursuper}) one should 
use a different gauge, e.g.~Kruskal gauge or Israel gauge~\cite{Israel:1966}. 
The construction of an atlas by means of ``large'' Eddington-Finkelstein 
patches and ``small'' Kruskal patches has been introduced 
by Walker~\cite{Walker:1970}. By tuning the matter contributions in 
a special way it may be possible to achieve $X^+=0=X^-$ in an 
open region which implies that also the dilaton field $X$ 
(and thus the surface area) has to be constant in that region. 
For minimal coupling to the dilaton it follows from~(\ref{eq:pf6}) 
that such a region has constant curvature and thus may be only 
Minkowski, Rindler or (A)dS. Note that in the absence of matter 
and cosmological constant it is not possible to achieve such a 
constant dilaton region for finite $X$. Thus, the appearance of 
such regions is a non-trivial consequence of the presence of matter.

\subsection{Quantisation of SSSMG}

There are two basic strategies: either to quantise first and to impose 
symmetries later or the other way round. 
The first one appears to be preferable conceptually, but it is 
more difficult to implement. 
Since one of the points of imposing spherical symmetry is 
to simplify the quantisation procedure itself it is also tempting 
to take the second route. At least some of the basic conceptual 
problems arise even in this simplified framework and provided 
they can be solved one can learn something for the full theory 
without introducing unnecessary technical difficulties. Actually, 
the preference for either of the two strategies depends on 
which kind of question one would like to ask; it is not just 
a matter of taste. We would like to be more concrete on this: 
First of all, one should recall that in both cases there are no 
propagating physical modes in the gravity sector. Thus, if one 
is interested e.g.~in scattering problems where virtual BHs 
may arise as intermediate states one has to add matter degrees 
of freedom if one would like to keep spherical symmetry, as there 
are no spherically symmetric gravitons. Adding matter is much 
simpler following the second route. On the other hand, if one 
is interested in questions that may be addressed without matter 
the first route seems to be the better one as it allows for 
slightly more structure in the geometrical sector than the 
more restrictive first one. 
As we are interested in interactions with SM 
fields and ensuing questions of information loss, scattering 
problems, virtual BH production etc.~we impose spherical 
symmetry first and quantise later. Thus, we take~(\ref{eq:5.1}) as 
our starting point and try to quantise this effective action 
in two dimensions. 

There are still two alternatives: either one fixes a geometric background 
before quantisation and applies methods from quantum field 
theory on a curved background (cf.~e.g.~\cite{Birrell:1982ix}), 
thus encountering the phenomenon of Hawking radiation, or one 
quantises geometry exactly first and applies perturbative methods 
in the matter sector afterwards (cf.~e.g.~sect.~7 of~\cite{Grumiller:2002nm}). 
As there exists an extensive amount of literature devoted 
to the first route (even in the more general case when 
spherical symmetry is absent) the focus will be on the 
second path. Along these lines the simple case with a 
single scalar field in the matter sector has been studied 
extensively~\cite{Kummer:1998zs,Bergamin:2004us}, 
for reviews cf.~\cite{Grumiller:2002nm,Grumiller:2004yq}.
The extension to the SSSMG is straightforward, albeit 
somewhat lengthy. Thus, we will merely present the algorithm 
and note especially where differences to the previous cases 
arise, rather than presenting all calculations in detail.

The first step, a Hamiltonian analysis including a discussion of constraints, 
their algebra and the construction of the BRST charge and 
the ghost sector fortunately essentially remains the same. 
The algorithm works as follows: as starting point declare the 
zero-component as ``time'' and introduce canonical coordinates 
$q=(\om_1,e^-_1,e^+_1,a_1,b_1,{\rm matter})$ with associated 
momenta $p=(X,X^+,X^-,y,z,{\rm matter})$ and 
$\bar{q}=(\om_0,e^-_0,e^+_0,a_0,b_0)$ producing 
primary first class constraints $\bar{p}\approx 0$. 
In addition, there will be the usual second class 
constraints from the fermions, which may be dealt 
with in the standard way, i.e., by introducing Dirac 
brackets. Then, take the Dirac bracket of the first 
class constraints with the Hamiltonian to calculate 
the secondary constraints denoted by $G$ (which are 
also first class). It is then noticed that the Hamiltonian 
is a sum over constraints, $H=\Sigma\bar{q}G$, as expected 
for a reparameterisation invariant theory \cite{Henneaux:1992}. No
ternary constraints arise. Next, one should consider the structure
functions arising in the Dirac algebra of first class constraints. 
They will enter the BRST charge, which may be constructed 
straightforwardly and does not receive any higher order 
ghost contributions, i.e., no quartic ghost terms arise. At 
least in the geometric sector no ordering ambiguities arise, 
as discussed in appendix B.2 of \cite{Bergamin:2004us}. 
A convenient gauge-fixing fermion is one that leads to ``temporal'' gauge
\begin{align}
      \label{eq:quant1}
      \om_0=0\,,\quad e^-_0=1\,,\quad e^+_0=0\,,\quad a_0=0\,,\quad b_0=0\,.
\end{align}
In the geometric sector this amounts to Eddington-Finkelstein gauge. 
Note that it is not possible to set all zero components to zero 
because this would amount to a singular metric. The 
choice~(\ref{eq:quant1}) exploits the maximum amount of
simplification and consequently 
the gauge fixed Hamiltonian simplifies drastically as most of 
the terms drop out. The most convenient order of path 
integrations seems to be the following one: all non-geometric 
gauge fields, their related auxiliary fields and the 
corresponding ghost sectors are integrated out exactly. 
Then, the remaining ghost sector is eliminated yielding 
some (contribution to the Faddeev-Popov-)determinant in the measure. As a 
next step eventual matter momenta are integrated out, if 
this can be performed exactly by linear or Gaussian path 
integration. The ensuing action will be linear in the 
remaining zweibein and connection components. Thus, path 
integration over geometry can be performed yielding 
functional $\de$-functions. They can be used to perform 
the integration over the auxiliary fields $X,X^\pm$, 
cancelling exactly the first contribution to the Faddeev-Popov-determinant
mentioned above. Because the functional $\de$-functions contain first 
derivatives acting on the auxiliary fields at this point 
homogeneous solutions arise which have to be fixed conveniently. 
In ordinary QFT often ``natural boundary conditions'' are 
invoked, but clearly they cannot be implemented for all 
fields as the metric must not vanish asymptotically. Instead, 
a very natural and simple condition is asymptotic flatness, 
which indeed fixes the relevant homogeneous contributions. 
Irrelevant contributions may be absorbed by fixing the 
scaling- and shift-ambiguity of the dilaton field.
The path integral measure for the final matter integrations 
can be adjusted in accordance with~\cite{Toms:1987sh}. 
The ensuing effective action will be nonlocal and non-polynomial 
in the matter fields. So at the end of this algorithm one 
has an effective theory depending solely on the propagating 
physical modes (scalar fields and fermions), but, to 
emphasise this again as it is very important, this theory 
is {\em nonlocal} and non-polynomial. Physically,
non-locality comes from integrating out exactly the 
gravitational self-energy of the fields.

Perturbation theory may be imposed upon this effective theory and 
Feynman rules may be derived. Since the theory is non-polynomial 
tree vertices with an arbitrary number of external legs will emerge. 
Moreover, these vertices will be non-local, in general. 
For the special case of the spherically reduced Einstein-massless-Klein-Gordon model explicit results may be found in ref.~\cite{Fischer:2001vz}.
These 
Feynman rules are then the basis of any phenomenological study. 
It is not very difficult to obtain them, but 
somewhat tedious. 

\section{Concluding remarks}
\label{se:con}

The three formalisms discussed (Cartan, GHP, metric) were used to
spherically reduce the Standard model of particle physics 
with gravity in a comprehensive manner yielding the SSSMG. 
One of the main aims of the present work was to link knowledge 
from particle physics on the one and gravity on the other hand. 

We are convinced that the SSSMG as a two-dimensional model
as presented in section~\ref{se:4} will be of use for further research. These
results can for example be used to study several classical aspects like
the existence of solitonic solutions, semi-classical aspects like the
Hawking radiation and the no-hair theorem and quantum aspects like the
role of the virtual black holes in scattering.

Moreover, we hope that our investigations lay the foundation
for the discussion of extensions of the spherically reduced Standard Model, like
(local) supersymmetry, non-metricity and the reduction of higher dimensional
models.



\lhead{ACKNOWLEDGEMENT}
\section*{Acknowledgement}

This work has been supported by projects P-14650-TPH and J-2330-N08 of the
Austrian Science Foundation (FWF) and by projects 622-1/2002, 798-1/2003
and 349-1/2004 of the Austrian Exchange Service (\"OAD) and
is part of the research project Nr. 01/04 
{\it Quantum Gravity, Cosmology and Categorification} 
of the Austrian Academy of Sciences (\"OAW) and the National
Academy of Sciences of Ukraine (NASU).
Part of this work has been performed in the hospitable 
atmosphere of the International Erwin Schr\"{o}dinger 
Institute.
\\ \mbox{} \\
We deeply thank W.~Kummer and D.~Vassilevich for useful discussions. 

\lhead{\leftmark}
\appendix

\section{Spherical harmonics}
\label{app_Y}
Spherical harmonics for $s=0$ and $j=1$ are
\begin{align}
     {}_{0}Y_{1\,0}=\frac{1}{2}\sqrt{\frac{3}{\pi}}\cos\negthinspace\theta\,,\qquad
     {}_{0}Y_{1\,\pm 1}=\mp\frac{1}{2}\sqrt{\frac{3}{2\pi}}
      \sin\negthinspace\theta e^{\pm i\phi}\,.
\end{align}
The spin weighted spherical harmonics are given for spin 
weight $s=\pm \frac{1}{2}$ and $j=\frac{1}{2}$. Since
$-j \leq m \leq j$ one only has $m=\pm\frac{1}{2}$. With
\begin{align}
        \overline{{}_s Y_{j,m}} = (-1)^{m+s} {}_{-s} Y_{j,-m},
        \label{bar_y}
\end{align}
one finds that there are only two independent spin
weighted spherical harmonics if $s=\pm \frac{1}{2}$, $j=\frac{1}{2}$
and $m=\pm\frac{1}{2}$. These are
\begin{align}
        {}_{\frac{1}{2}} Y_{\frac{1}{2}\, \frac{1}{2}} 
        &= \frac{1}{\sqrt{2 \pi}} \cos\frac{\theta}{2} e^{i \frac{\phi}{2}}, 
        \label{y1/2}\\
        {}_{-\frac{1}{2}} Y_{\frac{1}{2}\, \frac{1}{2}} 
        &=-\frac{1}{\sqrt{2 \pi}} \sin\frac{\theta}{2} e^{i \frac{\phi}{2}}.
        \label{y-1/2}
\end{align}
The other two spin weighted spherical harmonics
\begin{align}
        {}_{-\frac{1}{2}} Y_{\frac{1}{2}\, -\frac{1}{2}} 
        &= -\frac{1}{\sqrt{2 \pi}} \cos\frac{\theta}{2} e^{-i \frac{\phi}{2}}, 
        \\
        {}_{\frac{1}{2}} Y_{\frac{1}{2}\, -\frac{1}{2}} 
        &=-\frac{1}{\sqrt{2 \pi}} \sin\frac{\theta}{2} e^{-i \frac{\phi}{2}},
\end{align}
are obtained by using~(\ref{bar_y}).
One can easily check that these functions obey the orthogonality
condition~(\ref{ortho}).

\section{Linking GHP and Cartan formalism}
\label{app_link}

Using basis 1-forms $\bs{l}$, $\bs{n}$,
$\bs{m}$, $\bs{\bar{m}}$ the metric can be rewritten
to
\begin{align}
        \dif s^2 =2\, \bs{l} \otimes \bs{n} - 2\, \bs{m} \otimes \bs{\bar{m}}\,,
        \label{gab_form}
\end{align}
where the notation is the same as for the null tetrad~(\ref{nulltetrad}).
Note that the choice of the basis 1-forms is not
unique and can be changed in many more or less practical ways.
The spin coefficients can be read of from 
\begin{align}
        \dif \bs{l} &= \bs{m} \wedge \bs{l} \left(\beta' -\bar{\beta} +
 \bar{\tau} \right) + \bs{\bar{m}} \wedge \bs{l} \left(\bar{\beta}'
 -\beta + \tau \right) + \bs{l} \wedge \bs{n} \left(\gamma' +
 \bar{\gamma}' \right)  \nonumber \\
        &+ \bs{m} \wedge \bs{\bar{m}} \left(\rho -
 \bar{\rho}\right) + \bs{m} \wedge \bs{n} \ \bar{\kappa} + \bs{\bar{m}}
 \wedge \bs{n} \ \kappa, \label{cartan1} \\
        \dif \bs{m} &= \bs{m} \wedge \bs{l} \left(\bar{\gamma} - \gamma
 + \bar{\rho}' \right) + \bs{\bar{m}} \wedge \bs{l} \ \bar{\sigma}'
 \nonumber \\
        &+ \bs{m} \wedge \bs{n} \left(\gamma' - \bar{\gamma}'
 + \rho \right) + \bs{\bar{m}} \wedge \bs{n} \ \sigma
\nonumber \\ 
        &+ \bs{l} \wedge \bs{n} \left(\bar{\tau}' -\tau\right) + \bs{m} \wedge
 \bs{\bar{m}} \left(\beta + \bar{\beta}' \right). \label{cartan2}
\end{align}
 
\section{Inner products}
\label{innerproducts}
The notation is shortened by writing $\pm$ for $\pm 1/2$.
Furthermore the expansion coefficients $A,B,P,Q$ are left
out in the integrands because one can read them of by looking
at the spin weighted spherical harmonics.

Terms of type $(\bar{P} Q)(P \bar{Q})$ and 
$(\bar{A} B)(A \bar{B})$ for $s=\pm 1/2$ respectively read
\begin{align}
      \int 
      ({}_{\pm}\bar{Y}_{+ \pm}\, {}_{\pm}Y_{+ \pm})
      ({}_{\pm}Y_{+ \pm}\, {}_{\pm}\bar{Y}_{+ \pm})
      \dif \Omega^2 &= \frac{1}{3 \pi}\,, \\
      \int 
      ({}_{\pm}\bar{Y}_{+ \pm}\, {}_{\pm}Y_{+ \pm})
      ({}_{\pm}Y_{+ \mp}\, {}_{\pm}\bar{Y}_{+ \mp})
      \dif \Omega^2 &= \frac{1}{6 \pi}\,, \\
      \int 
      ({}_{\pm}\bar{Y}_{+ \pm}\, {}_{\pm}Y_{+ \mp})
      ({}_{\pm}Y_{+ \pm}\, {}_{\pm}\bar{Y}_{+ \mp})
      \dif \Omega^2 &= \frac{1}{6 \pi}\,.
\end{align}
Terms of type $(\bar{P} Q)(A \bar{B})$ are
\begin{align}
      \int 
      ({}_{+}\bar{Y}_{+ \pm}\, {}_{+}Y_{+ \pm})
      ({}_{-}Y_{+ \mp}\, {}_{-}\bar{Y}_{+ \mp})
      \dif \Omega^2 &= \frac{1}{3 \pi}\,, \\
      \int 
      ({}_{+}\bar{Y}_{+ \pm}\, {}_{+}Y_{+ \pm})
      ({}_{-}Y_{+ \pm}\, {}_{-}\bar{Y}_{+ \pm})
      \dif \Omega^2 &= \frac{1}{6 \pi}\,, \\ 
      \int 
      ({}_{+}\bar{Y}_{+ \pm}\, {}_{+}Y_{+ \mp})
      ({}_{-}Y_{+ \pm}\, {}_{-}\bar{Y}_{+ \mp})
      \dif \Omega^2 &= -\frac{1}{6 \pi}\,.
\end{align}
Terms of type $(\bar{A} B)(P \bar{Q})$ immediately
follow from complex conjugation. Thus on gets 
\begin{multline}
      \int \bar{P} Q P \bar{Q} \Phi^2 \dif \Omega^2 \omega_{g^{(2)}} =    
      \frac{1}{3 \pi} \int \Bigl(
      (\bar{P}_{+ \pm} Q_{+ \pm} P_{+ \pm} \bar{Q}_{+ \pm})
      \\
      + \frac{1}{2}(\bar{P}_{+ \pm} Q_{+ \pm} P_{+ \mp} \bar{Q}_{+ \mp}
      + \bar{P}_{+ \pm} Q_{+ \mp} P_{+ \pm} \bar{Q}_{+ \mp})
      \Bigr) \Phi^2  \omega_{g}\,,
      \label{app_inn1}
\end{multline}
\begin{multline}
      \int \bar{A} B A \bar{B} \Phi^2 \dif \Omega^2 \omega_{g^{(2)}} =    
      \frac{1}{3 \pi} \int \Bigl(
      (\bar{A}_{+ \pm} B_{+ \pm} A_{+ \pm} \bar{B}_{+ \pm})
      \\
      + \frac{1}{2}(\bar{A}_{+ \pm} B_{+ \pm} A_{+ \mp} \bar{B}_{+ \mp}
      + \bar{A}_{+ \pm} B_{+ \mp} A_{+ \pm} \bar{B}_{+ \mp})
      \Bigr) \Phi^2  \omega_{g}\,.
      \label{app_inn2}
\end{multline}
For the mixed ones one finds
\begin{multline}
      \int \bar{P} Q A \bar{B} \Phi^2 \dif \Omega^2 \omega_{g^{(2)}} =    
      \frac{1}{3 \pi} \int \Bigl(
      (\bar{P}_{+ \pm} Q_{+ \pm} A_{+ \mp} \bar{B}_{+ \mp})
      \\
      + \frac{1}{2}(\bar{P}_{+ \pm} Q_{+ \pm} A_{+ \pm} \bar{B}_{+ \pm}
      - \bar{P}_{+ \pm} Q_{+ \mp} A_{+ \pm} \bar{B}_{+ \mp})
      \Bigr) \Phi^2  \omega_{g}\,,
      \label{app_inn3}
\end{multline}
together with its complex conjugate.

The terms of the inner products can be written in terms 
of two-spinors
\begin{alignat}{2}
      &\bar{\Psi}^{\I}_{jm}P_{+}\Psi^{\I}_{jm}=\bar{B}_{jm}A_{jm}\,,&\qquad
      &\bar{\Psi}^{\I}_{jm}P_{-}\Psi^{\I}_{jm}=\bar{A}_{jm}B_{jm} \\
      &\bar{\Psi}^{\II}_{jm}P_{+}\Psi^{\II}_{jm}=-\bar{P}_{jm}Q_{jm}\,,&\qquad
      &\bar{\Psi}^{\II}_{jm}P_{-}\Psi^{\II}_{jm}=-\bar{Q}_{jm}P_{jm}\,.
\end{alignat}
Hence we find
\begin{multline}
      \int (\bar{A}BA\bar{B}+\bar{P}QP\bar{Q})\Phi^2 \dif\Omega^2 \omega_{g}=\\   
      \frac{1}{3 \pi}\sum_{m=\pm\frac{1}{2}}\int\Bigl(
      (\bar{\Psi}^{\III}_{\frac{1}{2}\,\pm m}P_{+}
      \Psi^{\III}_{\frac{1}{2}\,\pm m})
      (\bar{\Psi}^{\III}_{\frac{1}{2}\,\pm m}P_{-}
      \Psi^{\III}_{\frac{1}{2}\,\pm m})\phantom{xxxxxxxxxx}\\+
      \frac{1}{2}
      (\bar{\Psi}^{\III}_{\frac{1}{2}\,\mp m}P_{+}
      \Psi^{\III}_{\frac{1}{2}\,\mp m})
      (\bar{\Psi}^{\III}_{\frac{1}{2}\,\pm m}P_{-}
      \Psi^{\III}_{\frac{1}{2}\,\pm m})\\+
      \frac{1}{2}
      (\bar{\Psi}^{\III}_{\frac{1}{2}\,\mp m}P_{+}
      \Psi^{\III}_{\frac{1}{2}\,\pm m})
      (\bar{\Psi}^{\III}_{\frac{1}{2}\,\pm m}P_{-}
      \Psi^{\III}_{\frac{1}{2}\,\mp m})
      \Bigr)\Phi^2 \omega_{g}\,,
      \label{d1}
\end{multline}
where the first and second term of the left-hand side is obtained if 
$\Psi^{\I}$ or $\Psi^{\II}$ is considered respectively.

For the mixed terms we find
\begin{multline}
      \int (\bar{P}QA\bar{B}+\bar{Q}PB\bar{A})\Phi^2 \dif\Omega^2 \omega_{g}=\\
      -\frac{1}{3 \pi}\sum_{m=\pm\frac{1}{2}}\int\Bigl(
      (\bar{\Psi}^{\I}_{\frac{1}{2}\,\mp m}P_{\pm}
      \Psi^{\I}_{\frac{1}{2}\,\mp m})
      (\bar{\Psi}^{\II}_{\frac{1}{2}\,\pm m}P_{\pm}
      \Psi^{\II}_{\frac{1}{2}\,\pm m})\phantom{xxxxxxxxxx}\\+
      \frac{1}{2}
      (\bar{\Psi}^{\I}_{\frac{1}{2}\,\pm m}P_{\pm}
      \Psi^{\I}_{\frac{1}{2}\,\pm m})
      (\bar{\Psi}^{\II}_{\frac{1}{2}\,\pm m}P_{\pm}
      \Psi^{\II}_{\frac{1}{2}\,\pm m})\\-
      \frac{1}{2}
      (\bar{\Psi}^{\I}_{\frac{1}{2}\,\pm m}P_{\pm}
      \Psi^{\I}_{\frac{1}{2}\,\mp m})
      (\bar{\Psi}^{\II}_{\frac{1}{2}\,\pm m}P_{\pm}
      \Psi^{\II}_{\frac{1}{2}\,\mp m})
      \Bigr)\Phi^2 \omega_{g}\,,
      \label{d2}
\end{multline}     
where the upper sign of the projector corresponds to the first term
of the left-hand side and the lower to the second.


\lhead{REFERENCES}



\providecommand{\href}[2]{#2}\begingroup\raggedright\endgroup

\end{document}